\RequirePackage{lineno} 

\documentclass[twocolumn,showpacs,preprintnumbers,amsmath,amssymb]{revtex4}

\usepackage{graphicx}
\usepackage{dcolumn}
\usepackage{bm}

\usepackage{epstopdf}

\def\bea{\begin{eqnarray}}
\def\eea{\end{eqnarray}}

\def\pp{\mbox{$p$-$p$}}
\def\pa{\mbox{$p$-$A$}}
\def\da{\mbox{$d$-$A$}}
\def\auau{\mbox{Au-Au}}

\def\aa{\mbox{$A$-$A$}}
\def\nn{\mbox{$N$-$N$}}

\def\pt{$p_t$}
\def\yt{$y_t$}
\def\nch{$n_{ch}$}

\begin{document} 

\setlength{\pdfpagewidth}{8.5in}
\setlength{\pdfpageheight}{11in}

\setpagewiselinenumbers
\modulolinenumbers[5]

\preprint{Version 3.2}

\title{Charge-multiplicity dependence of single-particle transverse-rapidity $\bf y_t$ and pseudorapidity $\bf \eta$ densities and 2D angular correlations from 200 GeV p-p collisions
}

\author{Thomas A.\ Trainor}\affiliation{CENPA 354290, University of Washington, Seattle, Washington 98195}
\author{Duncan J.\ Prindle}\affiliation{CENPA 354290, University of Washington, Seattle, Washington 98195}


\date{\today}

\begin{abstract}
An established phenomenology and theoretical interpretation of $p$-$p$ collision data at lower collision energies should provide a reference for $p$-$p$ and other collision systems at higher energies, against which claims of novel physics may be tested. The description of $p$-$p$ collisions at the relativistic heavy ion collider (RHIC) has remained incomplete even as claims for collectivity and other novelties in data from smaller systems at the large hadron collider (LHC) have emerged recently. In this study we report the charge-multiplicity dependence of two-dimensional (2D) angular correlations and of single-particle (SP) densities on transverse rapidity $y_t$ and pseudorapidity $\eta$ from 200 GeV $p$-$p$ collisions. We define a comprehensive and self-consistent two-component (soft + hard) model (TCM) for hadron production and report a significant $p$-$p$ nonjet (NJ) quadrupole component as a third (angular-correlation) component. Our results have implications for $p$-$p$\ centrality, the underlying event (UE), collectivity in small systems and the existence of flows in high-energy nuclear collisions.
\end{abstract}


\maketitle

 \section{Introduction} \label{intro}

The hadronic final state of \pp\ collisions at 200 GeV may provide a reference for other high-energy nuclear collisions at the relativistic heavy ion collider (RHIC) and large hadron collider (LHC). Claims for novel physics at higher energies or in \pa, \da\ or \aa\ collision systems should be based on an accurate and self-consistent phenomenology for conventional \pp\ processes at 200 GeV. However, current theoretical and experimental descriptions of high-energy \pp\ collisions appear to be incomplete.
Several unresolved aspects of \pp\ collisions are notable: 
(a) the role of \pp\ collision centrality in relation to the low-$x$ gluon transverse structure of the proton~\cite{ppcent2}, 
(b) the nature and systematics of the \pp\ {\em underlying event} (UE) defined as complementary to contributions from an event-wise-triggered high-energy dijet~\cite{cdfue,cmsue}, 
(c) the systematics of {\em minimum-bias} (MB) dijet (minijet) production manifested in \pp\ spectra and angular correlations~\cite{ppprd,porter2,porter3} and 
(d) possible existence and phenomenology of a {\em nonjet azimuth quadrupole} component in \pp\ 2D angular correlations previously studied in \aa\ collisions (as quantity $v_2$)~\cite{davidhq,davidhq2,noelliptic,v2ptb}, especially in connection with a claimed same-side ridge observed in LHC \pp\ angular correlations~\cite{cmsridge}. A more detailed discussion of those issues is presented in Sec.~\ref{issues}.

In the present study we establish a more complete mathematical model for \pp\ phenomenology based on the  $n_{ch}$ dependence of single-particle (SP) $p_t$ spectra and \pt-integral $\eta$ densities and \pt-integral 2D angular correlations. We confront several issues: Is there any connection between $n_{ch}$ and \pp\ centrality? Is \pp\ centrality a relevant concept? A nonjet (NJ) quadrupole component in \aa\ collisions is the complement to jet-related and projectile-fragment correlations. Is there an equivalent phenomenon in \pp\ collisions, and what might a \pp\ NJ quadrupole component reveal about \pp\ centrality or UE structure?  The \pp\ phenomenological model should offer a conceptual context with two manifestations: (a) as a mathematical framework to represent \pp\ data systematics efficiently, and (b) as a  theoretical framework to provide physical interpretation of model elements via comparisons between data structures and QCD theory.

Preliminary responses to such questions were presented in Ref.~\cite{pptheory}. They are supplemented here by new \pp\ SP density and 2D angular-correlation measurements. We emphasize the $n_{ch}$ dependence of angular correlations from \pp\ collisions, extending the \pp\ two-component model (TCM) to include a NJ quadrupole component previously extrapolated from measurements in \auau\ collisions~\cite{davidhq,davidhq2,gluequad} and now obtained directly from \pp\ 2D angular correlations. We establish $n_{ch}$-dependent phenomenology for soft (projectile proton dissociation), hard (parton fragmentation to MB dijets) and NJ quadrupole components in a {\em three}-component model and explore possible correspondence with \pp\ centrality, UE structure, dijet production and the partonic structure of projectile protons. We also present a TCM for the \nch\ systematics of hadron densities on pseudorapidity $\eta$.

This article is arranged as follows:
Section~\ref{issues} summarizes open issues for \pp\ collisions.
Section~\ref{methods} reviews analysis methods for two-particle correlations.
Section~\ref{pptcm} describes a two-component model for hadron production in \pp\ collisions.
Section~\ref{ppangcorr} presents measured 2D angular correlations for 200 GeV \pp\ collisions.
Section~\ref{modelfits} summarizes the parametric results of 2D model fits to those \pp\ correlation data.
Section~\ref{jetcorr1} describes jet-related data systematics.
Section~\ref{njcorr} describes nonjet data systematics.
Section~\ref{etadensity} presents a two-component model for $\eta$ densities and the $\eta$-acceptance dependence of transverse-rapidity \yt\ spectra.
Section~\ref{syserr} discusses systematic uncertainties.
Section~\ref{ridgecms} reviews \pp\ same-side ``ridge'' properties and a proposed mechanism.
Sections~\ref{disc} and~\ref{summ}  present discussion and summary.

 \section{Open issues for $\bf p$-$\bf p$ collisions} \label{issues}

We present a summary of issues introduced in Sec. I including \pp\ centrality in relation to a conjectured underlying event, manifestations of MB dijets in spectra and correlations and existence and interpretation of a NJ quadrupole component of \pp\ 2D angular correlations.

\subsection{p-p centrality and the underlying event}

Item (a) of Sec.~\ref{intro} relates to interpretations of deep-inelastic scattering (DIS) data to indicate that low-$x$ gluons are concentrated within a transverse region of the proton substantially smaller than its overall size. It is argued that a high-$p_t$ dijet trigger may select more-central \pp\ collisions with greater soft-hadron production~\cite{ppcent2}. The soft (nonjet) multiplicity increase should be observed most clearly within a narrow azimuth {\em transverse region} (TR) centered at $\pi/2$ and thought to {\em exclude contributions from the triggered jets} centered at 0 and $\pi$.

Item (b) relates to measurements of charge multiplicity $N_\perp$ within the TR vs trigger condition $p_{t,trig}$ and $dN_\perp/dp_t$ spectra employed to characterize the UE~\cite{cdfue,cmsue}. Substantial increase of $N_\perp$ with higher $p_{t,trig}$ relative to a minimum-bias or non-single-diffractive (NSD) value is interpreted to reveal novel contributions to the UE, including {\em multiple parton interactions} (MPI) corresponding to a high rate of dijet production~\cite{mpi}. Monte Carlo \pp\ collision models such as PYTHIA~\cite{pythia} are tuned to accommodate such results~\cite{rick}. 

In Ref.~\cite{pptheory} items (a) and (b) were considered in the context of a two-component (soft+hard) model  (TCM) of hadron production as manifested in yields and spectra. It was observed that imposing a \pt\ trigger condition on events does lead to selection for {\em hard events} (containing at least one dijet) but that the soft component of the selected events is not significantly different from a MB event sample, in contrast to expectations from Ref.~\cite{ppcent2} that increased dijet frequency should correspond to more-central \pp\ collisions and therefore to a larger soft component from low-$x$ gluons. Since \nch\ apparently determines dijet rates directly~\cite{ppprd} it might also control \pp\ centrality, but Ref.~\cite{pptheory} concluded that further \pp\ correlation measurements are required to explore that possibility. The present study responds with the \nch\ dependence of MB dijet correlations and NJ quadrupole systematics that speak to the issue of \pp\ centrality and UE systematics.

\subsection{Manifestations of minimum-bias dijets}

Item (c) relates to the role of MB dijets in yields, spectra and various types of two-particle correlations. The contribution of MB dijets (minijets~\cite{minijets}) to SP \pt\ spectra was established in Refs.~\cite{ppprd,fragevo,hardspec}, and the contribution of minijets to 2D angular correlations was identified in Refs.~\cite{porter2,porter3,anomalous,jetspec}. However, further effort is required to establish a complete and self-consistent description of MB dijets in \pp\ and \aa\ yields, spectra and correlations.

In Ref.~\cite{anomalous} item (c) was addressed with 2D model fits applied to angular correlations from \auau\ collisions at 62 and 200 GeV to isolate several correlation components, including structures attributed to MB dijets and a NJ quadrupole, with emphasis on the former in that study. The systematics of two components (soft + hard) are consistent with the TCM. The dijet (hard-component) trend on centrality exhibits a {\em sharp transition} near 50\% fractional cross section below which \auau\ collisions appear to be simple linear superpositions of \nn\ binary collisions (transparency) and above which {\em quantitative} changes in the dijet component appear, but not in the NJ quadrupole component~\cite{noelliptic}. The MB dijet interpretation has been questioned variously, for more-central \aa\ collisions~\cite{glasma2} or for all nuclear collisions~\cite{nominijets}.

We wish to confirm the role of MB dijets as such via a self-consistent description of \auau\ {\em and} \pp\ collisions based on QCD theory.  Although a TCM for \pp\ yields and spectra vs \nch\ has been established~\cite{ppprd,fragevo} the systematics of MB dijet production in \pp\ collisions is incomplete. 
MB 2D angular correlations for \pp\ collisions have been decomposed into soft and hard components via a single \pt\ cut (at 0.5 GeV/c)~\cite{porter2}, but a TCM for \pp\ angular correlations vs \nch\ has not been available. In Ref.~\cite{jetspec} \auau\ MB jet-related correlation structure vs centrality was related quantitatively to spectrum hard components (dijets) to establish a direct link of both data formats with pQCD predictions. In the present study we carry out a similar analysis of \pp\ vs \nch\  trends.

We also extend the \pp\ TCM established on marginal \yt\  to the 2D $(y_t,\eta)$ system to determine the distribution of minijets on the full SP momentum space. The extension to $\eta$ may provide further evidence  that a MB dijet interpretation of  the inferred TCM hard component is {\em necessary} as a distinct element of hadron production.

\subsection{Nonjet azimuth quadrupole}

Item (d) relates to the possibility of a significant amplitude for a unique azimuth quadrupole in \pp\ collisions. (The NJ quadrupole should be distinguished from the quadrupole component of a {\em jet-related} 2D peak projected onto 1D azimuth.) Measurements of a NJ quadrupole component of angular correlations in \auau\ collisions (conventionally represented by parameter $v_2$) are found to be consistent with a simple universal trend on centrality and collision energy extrapolating to a nonzero value for \nn\ collisions~\cite{davidhq}. The extrapolation is consistent with a QCD-theory prediction for $v_2$ in \pp\ collisions~\cite{boris}. 

The \pp\ NJ quadrupole may be related to a same-side ``ridge'' reported in \pp\ collisions at 7 TeV (with special cuts on \pt\ and \nch\ imposed)~\cite{cms,cmsridge}. It has been suggested that the \pp\ same-side ridge arises from the same mechanism proposed for \aa\ collisions based on collective motion (flows) coupled to initial-state collision geometry. Systematics of a possible NJ quadrupole in \pp\ collisions have thus emerged as an important new topic.

Reference~\cite{cmsridge} considered extrapolation of NJ quadrupole centrality systematics in 200 GeV \auau\ collisions to \nn\ collisions, and further extrapolation to LHC energies based on measured RHIC energy dependence. In that scenario  the same-side ridge observed in LHC \pp\ collisions corresponds to one lobe of the NJ quadrupole. The other lobe is obscured by the presence of a dominant away-side (AS) 1D jet peak. Quantitative correspondence was observed in Ref.~\cite{cmsridge} suggesting that the NJ quadrupole may play a significant role in \pp\ collisions, but no direct \pp\ quadrupole measurements existed. This study offers a response to that issue.

Measurement of NJ quadrupole trends may shed light on the question of \pp\ centrality [item (a)] by analogy with \aa\ quadrupole systematics wherein the NJ quadrupole measured by a {\em per-particle} variable first increases rapidly with centrality and then falls sharply toward zero with decreasing \aa\ eccentricity, as described by a Glauber model  based on the eikonal approximation. Since dijet production vs \nch\ in \pp\ collisions suggests that the eikonal approximation is not valid for that system~\cite{tomalicempt,tomaliceptfluct} the NJ quadrupole trend could provide a critical test of the eikonal assumption for \pp\ collisions.

 \section{Analysis methods} \label{methods}

We review technical aspects of two-particle angular-correlation analysis methods applied to \pp\ collisions at the RHIC. Further method details appear in Refs.~\cite{ppprd,hardspec,porter1,porter2,porter3,inverse,axialci,axialcd,anomalous,davidhq,davidhq2}.





\subsection{Kinematic measures and spaces}

High-energy nuclear collisions produce final-state hadrons as a distribution within cylindrical 3D momentum space $(p_t,\eta,\phi)$, where $p_t$ is transverse momentum, $\eta$ is pseudorapidity and $\phi$ is azimuth angle. Transverse mass is $m_t = \sqrt{p_t^2 + m_h^2}$ with hadron mass $m_h$. Pseudorapidity is  $\eta = -\ln[\tan(\theta/2)] $ ($\theta$ is polar angle relative to collision axis $z$), and $\eta \approx \cos(\theta)$ near $\eta = 0$. To improve visual access to low-$p_t$ structure and simplify description of the \pt-spectrum hard component (defined below) we present spectra on transverse rapidity $y_t = \ln[(m_t + p_t) / m_h]$.  For unidentified hadrons $y_t$, with pion mass assumed (about 80\% of hadrons), serves as a regularized logarithmic $p_t$ measure. 
A typical detector acceptance $p_t > 0.15$ GeV/c corresponds to $y_t > 1$.

Correlations are observed in two-particle momentum space $(p_{t1},\eta_1,\phi_1,p_{t2},\eta_2,\phi_2)$.
An {\em autocorrelation} on angular subspace $(x_1,x_2)$ (where $x = \eta$ or $\phi$) is derived by averaging pair density $\rho(x_1,x_2)$ along diagonals on $(x_1,x_2)$ parallel to the sum axis $x_\Sigma = x_1 + x_2$~\cite{inverse}. The averaged pair density $\rho(x_\Delta)$ on defined {\em difference variable} $x_\Delta = x_1 - x_2$ is then an autocorrelation. The notation $x_\Delta$ rather than $\Delta x$ for difference variables is adopted to conform with mathematical notation conventions and to retain $\Delta x$ as a measure of a detector acceptance on parameter $x$. 
For correlation structure approximately independent of $x_\Sigma$ over some limited acceptance $\Delta x$ (stationarity, typical over $2\pi$ azimuth and within some limited pseudorapidity acceptance $\Delta \eta$) angular correlations remain undistorted (no information is lost in the projection by averaging).  
\pt-integral 2D angular autocorrelations are thus lossless projections of 6D two-particle momentum space onto angle difference axes $(\eta_\Delta,\phi_\Delta)$. 
The $\phi_\Delta$ axis is divided into {\em same-side} (SS, $|\phi_\Delta| < \pi/2$) and {\em away-side}  (AS, $\pi/2 < |\phi_\Delta| < \pi$) intervals.

\subsection{p-p initial-state geometry}

For collisions between two composite projectiles the collision final state (FS) may depend on the transverse separation of the collision partners (impact parameter $b$) and the phase-space distribution of constituents within each projectile, collectively the initial-state (IS) geometry. We wish to determine how the IS geometry relates to an observable derived from FS hadrons and how the IS influences FS hadron yields, spectra and correlations.

For A-A collisions the projectile constituents are nucleons $N$ all sharing a common lab velocity (modulo Fermi motion) and distributed over a nuclear volume. Based on a Glauber model of \aa\ collisions (assuming the eikonal approximation) nucleons are classified as participants (total number $N_{part}$) or spectators, and the mean number of \nn\ binary encounters $N_{bin}$ is estimated. The relation $N_{bin} \sim N_{part}^{4/3}$ is a consequence of the eikonal approximation. Parameters $N_{part}$ and $N_{bin}$, depending on \aa\ impact parameter $b$, are in turn related to macroscopic FS observable \nch\ within some angular acceptance via the MB cross-section distribution $d\sigma(b) / dn_{ch}$.

For \pp\ collisions the constituents are partons distributed on the transverse configuration space of projectile protons {\em and} on longitudinal-momentum fraction $x$ (fraction of proton momentum carried by a parton). One could apply a similar Glauber approach to \pp\ IS geometry, including assumed eikonal approximation as in the \aa\ description (e.g.\ default PYTHIA~\cite{pythia}). As noted in the introduction, it is conjectured that imposing a dijet trigger should favor more-central \pp\ collisions and therefore a substantial increase in soft-hadron production from low-$x$ gluons~\cite{ppcent2}.
However, some aspects of \pp\ collision data appear to be inconsistent with such a description, specifically parton transverse position and a \pp\ impact parameter. Nevertheless, FS measures for number of participant low-$x$ partons and their binary encounters may be relevant and experimentally accessible~\cite{pptheory}.

\subsection{p-p $\bf y_t \times y_t$ and 2D angular correlations}

$\rho(\vec p_{1},\vec p_{2})$ represents a basic pair density on 6D pair momentum space.  The event-ensemble-averaged pair density  $\bar \rho_{sib}$ derived from sibling pairs (pairs drawn from single events) includes the correlation structures to be measured. $\bar \rho_{mix}$ is a density of mixed pairs drawn from different but similar events. $\rho_{ref}$ denotes a minimally-correlated reference-pair density derived from (a) a mixed-pair density or (b) a Cartesian product of SP angular densities $\bar \rho_0$ via a factorization assumption. 


Differential correlation structure is determined by comparing a sibling-pair density to a reference-pair density in the form of difference $\Delta \rho = \bar \rho_{sib} - \rho_{ref}$ representing a correlated-pair density or {\em covariance} density.  {\em Per-particle} measure $\Delta \rho / \sqrt{\rho_{ref}}$ has the form of Pearson's normalized covariance~\cite{pearson} wherein the numerator is a covariance and the denominator is approximately the geometric mean of marginal variances. In the Poisson limit a marginal variance may correspond to $\sigma^2_{n_{ch}} \approx  n_{ch} \propto \bar \rho_0$. Since $\rho_{ref} \approx \bar \rho_0 \times \bar \rho_0$ it follows that the geometric mean of variances is given by  $\sqrt{\rho_\text{ref}} \approx \bar \rho_0$ and the normalized covariance density is  a {per-particle} measure~\cite{inverse,ptscale,ptedep}. The number of final-state charged hadrons $\bar \rho_0 \sim n_{ch}$ in the denominator can be seen as a place holder. Other particle degrees of freedom may be more appropriate for various physical mechanisms (e.g.\ number of participant nucleons in \aa\ collisions, number of participant low-$x$ partons in \pp\ collisions) as described below.

We define $\Delta \rho / \sqrt{\rho_{ref}} \equiv \bar \rho_0 (r-1)$ where pair ratio $r \equiv  \bar \rho_{sib} / \bar \rho_{mix}$ cancels instrumental effects. That per-particle measure is not based on a physical model. In some analyses a correlation amplitude is defined as $\bar N (r-1)$ with $\bar N \equiv \Delta \phi \Delta \eta \, \bar \rho_0$~\cite{cms}, but such an amplitude then relies on a specific detector acceptance, is not ``portable.'' 

To assess the relation of data to IS geometry we convert per-charged-hadron model-fit results to
 $(2/N_{part}) \Delta \rho = (2/N_{part})  [\bar \rho_0^2(r-1)]$, the quantity in square brackets representing the {\em number of correlated pairs} within the detector acceptance. For this \pp\ analysis we assume the soft-component density $\bar \rho_s \equiv n_s/\Delta \eta$ is an estimator for \pp\ IS $N_{part}/2$ (low-$x$ parton participants). Given the simplified notation $\Delta \rho  / \sqrt{\rho_{ref}} \rightarrow A$ we plot $(\bar \rho_0/\bar \rho_s)A$ to convert ``per-particle'' from FS hadrons to IS low-$x$ partons and obtain a more interpretable \pp\ per-particle measure.

Correlations on two-particle momentum space $(\vec p_1,\vec p_2)$ can be factorized into distributions on 2D transverse-momentum space $(p_{t1},p_{t2})$ or transverse-rapidity space $(y_{t1},y_{t2}) \leftrightarrow y_t \times y_t$~\cite{ytxyt} and on 4D angle space $(\eta_1,\phi_2,\eta_2,\phi_2)$ reducible with negligible information loss to autocorrelations on difference variables $(\eta_\Delta,\phi_\Delta)$~\cite{inverse,axialci,anomalous}. In this study we focus on minimum-bias ($y_t$-integral) 2D angular correlations. 
Each of the several features appearing in \pp\ 2D angular correlations (a correlation {\em component}) can be modeled within acceptance $\Delta \eta$ by a simple functional form  (a model {\em element}), including 1D and 2D Gaussians and azimuth sinusoids uniform on $\eta_\Delta$. The cosine elements $\cos(m\phi_\Delta)$ represent {\em cylindrical multipoles} with pole number $2m$, e.g., dipole, quadrupole and sextupole for $m = 1,~2,~3$. 
Angular correlations can be formed separately for like-sign (LS) and unlike-sign (US) charge combinations, as well as for charge-independent (CI = LS + US) and charge-dependent (CD = LS $-$ US) combinations~\cite{axialci,axialcd,porter2,porter3}. 

\subsection{p-p nonjet azimuth quadrupole}

The azimuth quadrupole ($m = 2$ Fourier) component is a prominent feature of \aa\ angular correlations, represented there by symbol $v_2 = \langle \cos(2\phi) \rangle$. The mean value is nominally relative to an estimated \aa\ reaction plane~\cite{2004}. $v_2$ data are conventionally interpreted to represent elliptic flow, a hydrodynamic (hydro) response to IS asymmetry in non-central \aa\ collisions~\cite{hydro}.

If 2D angular correlations are projected onto 1D azimuth {\em any} resulting distribution can be expressed exactly in terms of a Fourier series. The density of correlated pairs is then
\bea \label{nf}
\Delta \rho(\phi_\Delta) \hspace{-.06in} &=& \hspace{-.06in} \bar \rho_{sib} - \rho_{ref} = \Delta V_0^2 +   2\sum_{m=1}^\infty \hspace{-.02in} V_m^2 \cos(m\phi_\Delta),
\eea
defining the {\em power-spectrum} elements $V_m^2$  of autocorrelation density $\Delta \rho(\phi_\Delta)$.
The corresponding {\em per-pair} correlation measure is the ratio
\bea 
\frac{\Delta \rho}{\rho_{ref}} &=&  \Delta v_0^2 +  2\sum_{m=1}^\infty  v_m^2 \cos(m\phi_\Delta).
\eea

Some Fourier amplitudes from analysis of 1D azimuth projections may include contributions from more than one mechanism. For example, $v_2$ data from conventional 1D analysis may include contributions from jet-related (``nonflow'') as well as NJ (``flow'') mechanisms~\cite{noelliptic,v2ptb}.  In contrast, a complete model of 2D angular correlations {\em with $\eta_\Delta$-dependent elements} permits isolation of several production mechanisms including a NJ quadrupole component~\cite{davidhq,davidhq2}.  Fourier components from 2D correlation analysis are denoted by $V_m^2\{2D\}$ or $v_m^2\{2D\}$, and only the $m = 1$ and 2 (dipole and quadrupole) Fourier terms are {\em required} by 2D data histograms (see Sec.~\ref{ppangcorr})~\cite{anomalous}. For measure $\Delta \rho / \sqrt{\rho_{ref}}$ data derived from model fits to 2D angular correlations the quadrupole component is denoted by $A_Q\{2D\} \equiv V_2^2\{2D\} / \bar \rho_0 = \bar \rho_0 v_2^2\{2D\}$ since the factorized reference density is $\rho_{ref} \rightarrow \bar \rho^2_0$.

In the present study of \pp\ 2D angular correlations we admit the possibility that a significant NJ quadrupole component may persist in high-energy \pp\ collisions (not necessarily of hydro origin) and retain the corresponding model element in the 2D data model function Eq.~(\ref{modelfunc}). 

The measured \pt-integral NJ quadrupole data for \auau\ collisions are represented above 13 GeV by~\cite{davidhq}
\bea \label{loglog}
A_Q\{\text{2D}\}(b,\sqrt{s_{NN}}) &\equiv& \bar \rho_0(b) v_2^2\{\text{2D}\}(b,\sqrt{s_{NN}}) \\ \nonumber
&=& C R(\sqrt{s_{NN}}) N_{bin}(b) \epsilon_{2,opt}^2(b),
\eea
where $C = 4.5\pm 0.2 \times 10^{-3}$, the energy-dependence factor is $R(\sqrt{s_{NN}}) = \log(\sqrt{s_{NN}} / \text{13.5 GeV}) / \log(200 / 13.5)$, $N_{bin}$ is the estimated number of \nn\ binary encounters in the Glauber model, and $\epsilon_{2,opt}(b)$ is the \auau\ $m = 2$ overlap eccentricity assuming a continuous (optical-model) nuclear-matter distribution.
Equation~(\ref{loglog}) describes measured $y_t$-integral azimuth quadrupole data in heavy ion collisions for all centralities down to \nn\ collisions and energies above $\sqrt{s_{NN}} \approx 13$ GeV and represents factorization of energy and centrality dependence for the NJ quadrupole. The 2D quadrupole data are also consistent with $V_2^2\{\text{2D}\} = \bar \rho_0 A_Q\{\text{2D}\} \propto N_{{part}} N_{bin}\epsilon_{2,opt}^2$~\cite{noelliptic}, a centrality  trend that, modulo the IS eccentricity, {\em increases much faster than the dijet production rate}. 

A non-zero value $v_2 \approx 0.02$ from Eq.~(\ref{loglog}) extrapolated to \nn\ collisions agrees with a \pp\ QCD color-dipole prediction~\cite{boris}.  As one aspect of the present \pp\ correlation study we confirm extrapolation of the \auau\ NJ quadrupole centrality trend to \nn\  collisions and determine  the \nch\ dependence of \pp\ $A_Q\{2D\}$. \pp\ NJ quadrupole systematics may help clarify the concept of \pp\ centrality: Is an IS eccentricity relevant for \pp\ collisions; if so how does it vary with \nch?

\section{TCM for $\bf p$-$\bf p$ collisions}  \label{pptcm}

The two-component (soft+hard) model (TCM) of hadron production in high energy nuclear collisions has been reviewed in Refs.~\cite{ppprd,pptheory} for \pp\ collisions and Refs.~\cite{hardspec,fragevo,anomalous} for \aa\ collisions. The TCM serves first as a mathematical framework for data description and then, after comparisons with theory, as a basis for physical interpretation of data systematics.  The TCM has been interpreted to represent two main sources of final-state hadrons: longitudinal projectile-nucleon dissociation (soft) and large-angle-scattered (transverse) parton fragmentation (hard). In \aa\ collisions the two processes scale respectively proportional to $N_{part}$ (participant nucleons $N$) and $N_{bin}$ (\nn\ binary encounters). Analogous scalings for \pp\ collisions were considered in Ref.~\cite{pptheory}. 

The (soft + hard) TCM accurately describes most FS hadron yield and spectrum systematics~\cite{ppprd,hardspec,anomalous}, whereas there is no significant manifestation of the NJ quadrupole in yields and spectra~\cite{quadspec,noelliptic,v2ptb}. In contrast, the NJ quadrupole plays a major role in \aa\ 2D angular correlations and is measurable as such even for \pp\ collisions (per this study).  The TCM previously applied to yields and spectra must therefore be extended to include the NJ quadrupole as a third component of all high-energy nuclear collisions.  An effective TCM should be complete and self-consistent, capable of describing all aspects of data from any collision system.

\subsection{p-p single-particle $\bf y_t$ spectra} \label{ppspec1}

The joint single-charged-particle 2D (azimuth integral) density on $y_t$ and $\eta$ is denoted by  $\rho_0(y_t,\eta) = d^2 n_{ch}/ y_t dy_t d\eta$. The $\eta$-averaged (over $\Delta \eta$) \yt\ spectrum is $\bar \rho_0(y_t;\Delta \eta)$. The $y_t$-integral mean angular density is $\bar \rho_0(\Delta \eta) = \int dy_t y_t \bar \rho_0(y_t;\Delta \eta) \approx n_{ch} / \Delta \eta$  averaged over acceptance $\Delta \eta$ ($\eta$ averages are considered in more detail in Sec.~\ref{etadensity}). The $n_{ch}$ dependence of \pp\ $y_t$ spectra was determined in Ref.~\cite{ppprd}, and MB spectra were decomposed into soft and hard components according to the TCM.  

In \pp\ collisions soft and hard spectrum components have fixed forms but their relative admixture varies with $n_{ch}$~\cite{ppprd}.  The relation of the hard component to pQCD theory was established in Ref.~\cite{fragevo}. In \auau\ collisions the soft component retains its fixed form but the hard-component form changes substantially with centrality, reflecting quantitative modification of jet formation~\cite{fragevo}.  Identification of the hard component with jets in \pp\ and more-peripheral \auau\ collisions is supported by data systematics and comparisons with pQCD theory~\cite{fragevo,jetspec}. In more-central \auau\ collisions a jet interpretation for the TCM hard component has been questioned~\cite{anomalous}.

The two-component decomposition of \pp\ $y_t$ spectra conditional on uncorrected $n_{ch}'$ integrated over angular acceptance $\Delta \eta$ within $2\pi$ azimuth is denoted by~\cite{ppprd}
\bea \label{ppspec}
\bar \rho_0(y_t;n_{ch}') &=&  S(y_t;n_{ch}') + H(y_t;n_{ch}')
\\ \nonumber
&=& \bar \rho_s(n_{ch}') \hat S_0(y_t)  +  \bar \rho_{h}(n_{ch}') \hat H_0(y_t),
\eea
where  $\bar \rho_s = n_s / \Delta \eta$ and $\bar \rho_h = n_h / \Delta \eta$ are corresponding $\eta$-averaged soft and hard components with corrected $n_{ch} = n_s + n_h$ (see Sec.~\ref{etadensity}). The inferred soft and hard $y_t$ spectrum shapes [unit normal $\hat S_0(y_t)$ and $\hat H_0(y_t)$] are independent of $n_{ch}'$ and are just as defined in Refs.~\cite{hardspec,fragevo}.  The fixed hard-component spectrum shape $\hat H_0(y_t)$ (Gaussian plus power-law tail) is predicted quantitatively by measured fragmentation functions convoluted with a measured 200 GeV minimum-bias jet spectrum~\cite{fragevo,jetspec2}.

\begin{figure}[h]
  \includegraphics[width=1.65in,height=1.6in]{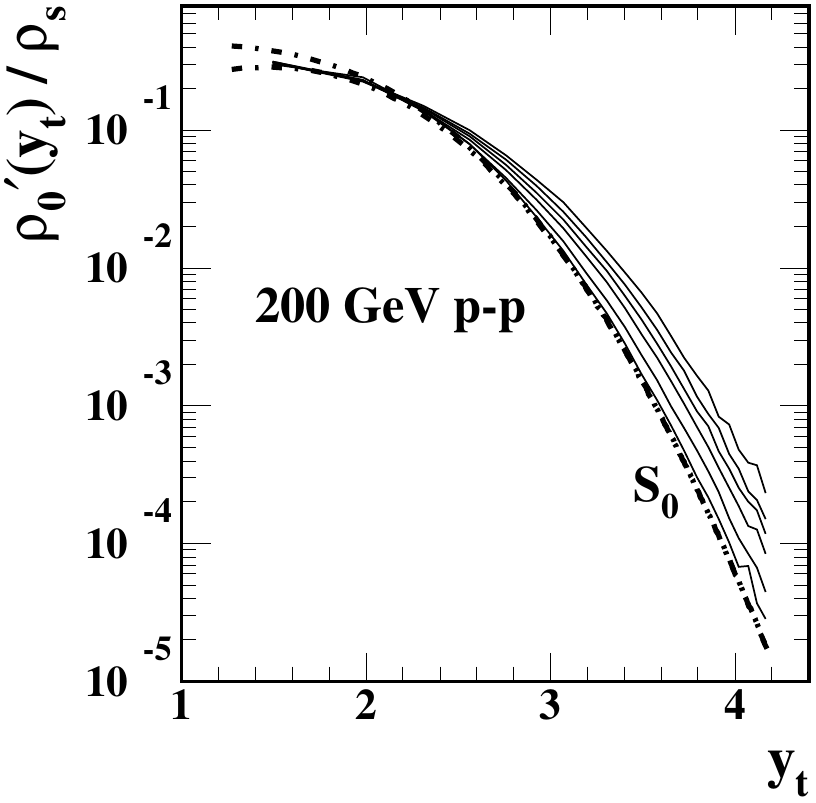}
  \includegraphics[width=1.65in,height=1.6in]{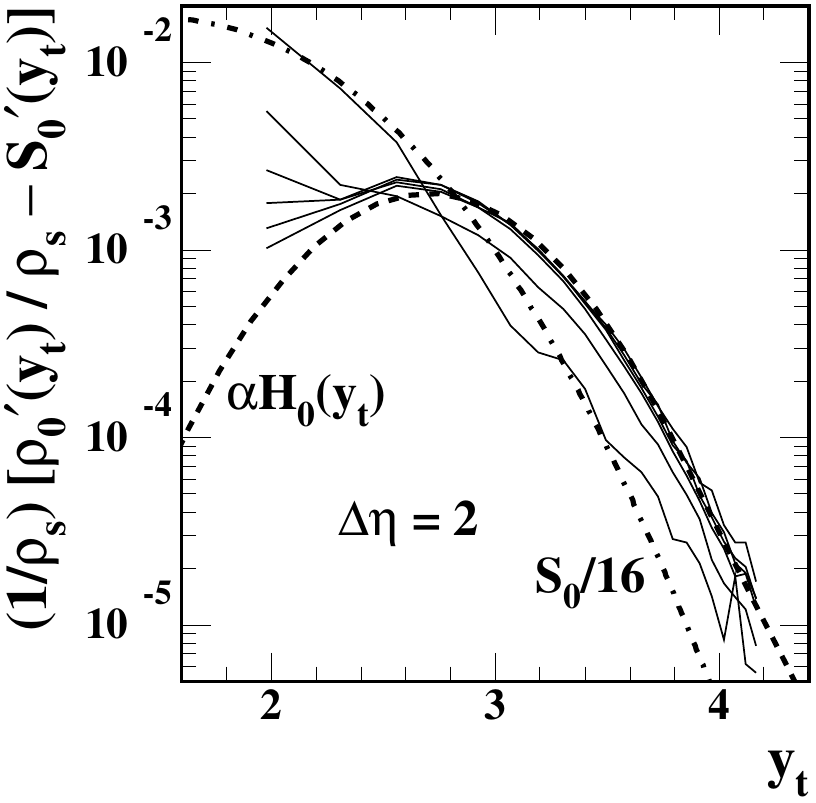}
\caption{\label{fig1a}
Left: Normalized \yt\ spectra for six multiplicity classes of 200 GeV \pp\ collisions ($n=1,\ldots,6$ see Table~\ref{multclass}). $\hat S_0(y_t)$ is the soft-component model function for corrected (upper dash-dotted) and uncorrected (lower dash-dotted) data. $\bar \rho_s$ is the soft-component multiplicity assuming $\alpha = 0.006$ (see text). The spectra are averaged over acceptance $\Delta \eta = 2$.
Right: Spectrum hard components in the form $H(y_t) / \bar \rho_s^2$ from Eq.~(\ref{ppspec}) compared to hard-component model function $\alpha \hat H_0(y_t)$ (dashed). Bars and carets are omitted from the figure labels.
 }  
\end{figure}

Figure \ref{fig1a} (left) shows $y_t$ spectra for several $n_{ch}'$ classes. The spectra (uncorrected for tracking inefficiencies) are normalized by corrected-yield soft component $\bar \rho_s$. A common $y_t$-dependent inefficiency function is introduced for comparison of this analysis with corrected spectra in Ref.~\cite{ppprd}, indicated below $y_t = 2$ by the ratio of the two dash-dotted curves representing uncorrected $S_0'(y_t)$ and corrected $\hat S_0(y_t)$ soft-component models. The data spectra are represented by spline curves rather than individual points to emphasize systematic variation with $n_{ch}'$.

Figure \ref{fig1a} (right) shows normalized spectra from the left panel in  the form $[\bar \rho_0'(y_t) / \bar \rho_s - S_0'(y_t)] / \bar \rho_s \approx \alpha \hat H_0(y_t)$ with $\alpha \approx 0.006$. From  Ref.~\cite{ppprd} we infer $\bar \rho_h \approx \alpha \bar \rho_s^2 = \alpha' \bar \rho_s'^2$. Given  that empirical relation and $n_{ch}' / \Delta \eta \equiv \bar \rho_0' = \bar \rho_s' + \bar \rho_h$ as simultaneous equations we can obtain $\bar \rho_s'$, $\bar \rho_s$ and $\bar \rho_h$ for any $n_{ch}'$ and $\Delta \eta$ (see details in Sec.~\ref{etayt}). Note that although the data hard-component shapes for the lowest two $n_{ch}'$ values deviate significantly from the $\alpha \hat H_0(y_t)$ model the integrals on \yt\ remain close to the value $\alpha$. The amplitude 0.33 of unit-normal $\hat H_0$ corresponds to the maximum of  the dashed curve  $0.006 \times 0.33 = 0.002$. These spectrum results are consistent with those from Ref.~\cite{ppprd} with $\Delta \eta = 1$ (see Sec.~\ref{etadep}).



\subsection{p-p dijet production} \label{ppdijet}

In Ref.~\cite{ppprd} the TCM spectrum hard-component yield $n_h$ within $\Delta \eta = 1$ was observed to vary as $n_h  \approx 0.01  n_{ch}' n_s$, with uncorrected $n_{ch}' \approx n_{ch}/2$.  A refined analysis provided the more precise density relation $\bar \rho_h \approx \alpha \bar \rho_s^2$. As noted, the TCM soft-component density $\bar \rho_s$ is then defined by the simultaneous equations $\bar \rho_0 = \bar \rho_s + \bar \rho_h$, and $\bar \rho_h = \alpha \bar \rho_s^2$ for some $\alpha = O(0.01)$, consistent with pQCD plus Ref.~\cite{ppprd}. In Ref.~\cite{pptheory}  $\bar \rho_s$ is associated with the number of {\em participant low-$x$ partons} (gluons) and dijet production, proportional to the number of participant-parton binary encounters, is then observed to scale $\propto \bar \rho_s^2$. 

Based on a dijet interpretation for the hard component we define $\bar \rho_h = n_h / \Delta \eta \equiv \epsilon(\Delta \eta) f(n_{ch}') 2\bar  n_{ch,j}$, where $f(n_{ch}')$ is the dijet frequency per collision and per unit $\eta$, $\epsilon(\Delta \eta) \in [0.5,1]$ is the average fraction of a dijet appearing in acceptance $\Delta \eta$, and $2\bar  n_{ch,j}$ is the MB mean dijet fragment multiplicity in $4\pi$. Dijet fraction $\epsilon(\Delta \eta)$ should be distinguished from initial-state eccentricity $\epsilon_{2,opt}$ associated with the NJ quadrupole. We also distinguish among number of dijets, number of jets, dijet mean fragment multiplicity and jet mean fragment multiplicity (their values integrated over $4\pi$ vs within some limited detector acceptance $\Delta \eta$). The definition of $\bar \rho_h$ above separates factors $\epsilon \in [0.5,1]$ and $f$ that were combined in previous studies. The $f$ values estimated here are thus approximately a factor 2 (i.e.\ $\approx 1/\epsilon$) larger than previous estimates~\cite{fragevo,jetspec}. 

For 200 GeV NSD \pp\ collisions with $ \bar \rho_s \approx 2.5$ and dijet mean fragment multiplicity $2\bar n_{ch,j} \approx 2.5 \pm 0.5$ derived from measured jet systematics the inferred jet frequency $f_{NSD} = 0.006\times  2.5^2 /(0.55 \times 2.5) \approx 0.027$ is inferred from \pp\ spectrum data within $\Delta \eta = 1$. That value can be compared with the pQCD prediction $f_{NSD} = \sigma_{dijet} / (\sigma_{NSD} \Delta \eta_{4\pi}) \approx 4.5~\text{mb} / (36.5~\text{mb} \times 5) \approx 0.025 $ for 200 GeV \pp\ collisions~\cite{fragevo} based on a measured jet spectrum and NSD cross section corresponding to mean-value \pp\ parton distribution functions. Thus, the observed NSD \pp\ spectrum hard-component yield $n_h$~\cite{ppprd} is quantitatively consistent with measured dijet systematics derived from event-wise jet reconstruction~\cite{ua1,cdfjets,jetspec2}.

If a non-NSD \pp\ event sample with arbitrary mean $n_{ch}'$ is selected we employ empirical \nch\ trends consistent with Ref.~\cite{ppprd} and having their own pQCD implications, as discussed in Ref.~\cite{pptheory}.  
We assume for the present study that the dijet frequency varies with soft multiplicity as
\bea \label{nj1}
f(n_{ch}') &\approx& 0.027  \left[\frac{\bar \rho_{s}(n_{ch}')}{\bar \rho_{s,NSD}}\right]^2
\eea
with $\bar \rho_{s,NSD} = 2.5$ for 200 GeV \pp\ collisions. We define $n_j(n_{ch}') = \Delta \eta\, f(n_{ch}')$ as the {\em dijet number} within some angular acceptance $\Delta \eta$. For the $n_{ch}'$ range considered in this study  the fraction of hard hadrons $n_h/n_{ch}$ is not more than about 15\%. The \pp\ final state is never dominated by hard processes but MB dijet production does play a major role, especially for $y_t < 3.3$ ($p_t < 2$ GeV/c) where {\em most jet fragments appear}.

\subsection{p-p two-particle correlations}

MB two-particle correlations have been studied extensively for NSD \pp\ collisions~\cite{porter1,porter2,porter3} and \auau\ collisions~\cite{axialci,anomalous,ytxyt}. A correspondence between jet-related correlations and $y_t$-spectrum hard components has been established quantitatively in Refs.~\cite{porter2,porter3,jetspec}. \pp\ angular-correlation structure is consistent with extrapolation of the centrality systematics of angular correlations from \auau\ collisions~\cite{anomalous}. Both correlations on transverse rapidity $y_t \times y_t$ and 2D angular correlations on $(\eta_\Delta,\phi_\Delta)$ from \pp\ collisions are described  by the TCM. 
$y_t \times y_t$ correlations for 200 GeV \pp\ collisions are fully consistent with the SP \yt\ spectrum results described above and in Ref.~\cite{ppprd}. The hard component corresponds (when projected onto 1D $y_t$) to the hard component of Eq.~(\ref{ppspec}). 

\begin{figure}[h]
  \includegraphics[width=1.65in,height=1.4in]{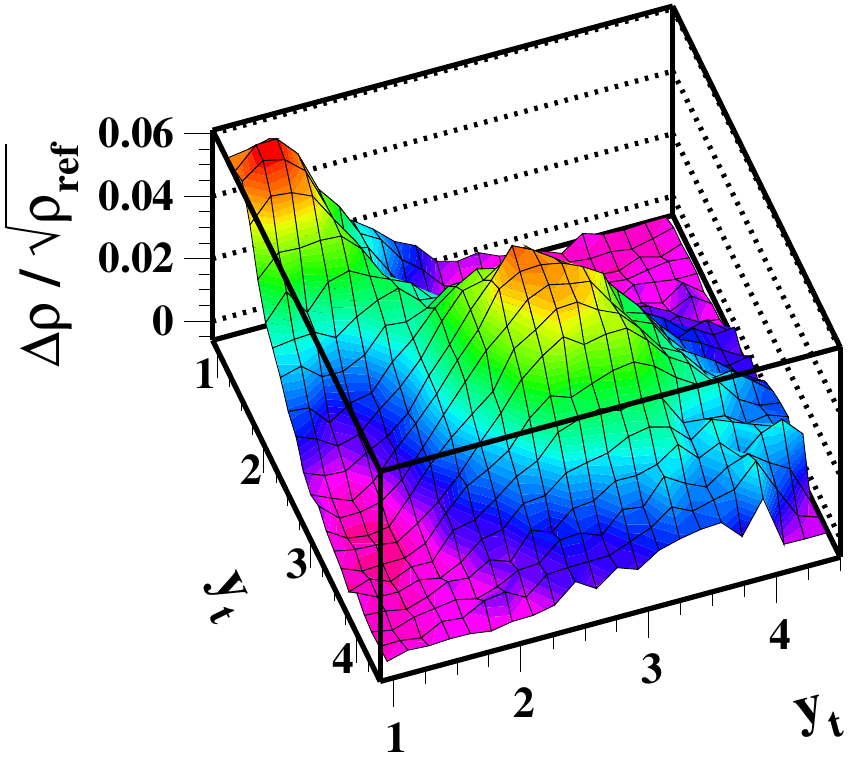}
 \put(-85,92) {\bf (a)}  
 \includegraphics[width=1.65in,height=1.4in]{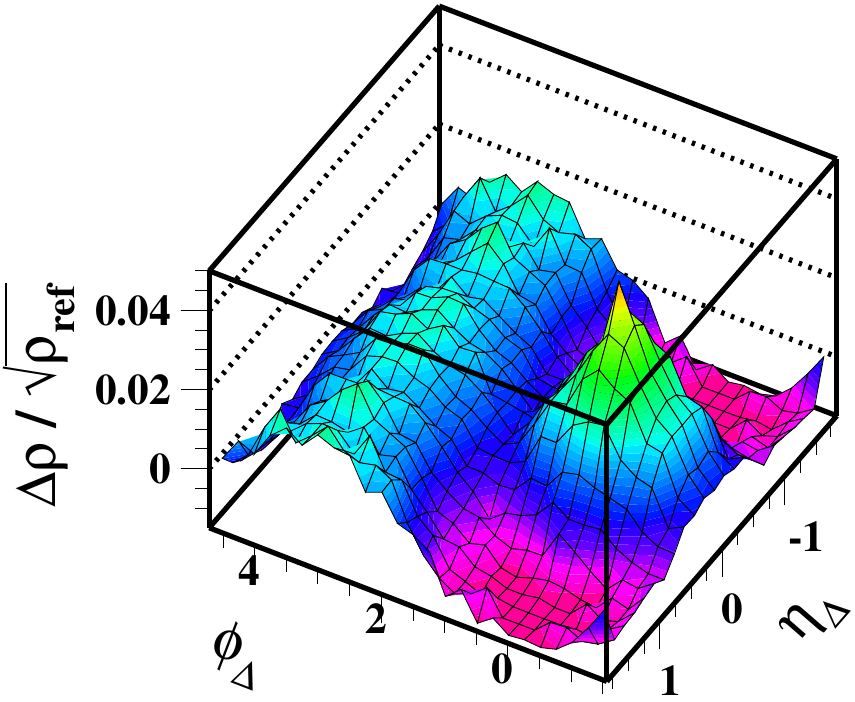}
 \put(-85,92) {\bf (b)}
 \caption{\label{ppcorr} (Color online)
(a) Minimum-bias correlated-pair density on 2D transverse-rapidity space $y_t \times y_t$ from 200 GeV \pp\ collisions showing soft (smaller \yt) and hard (larger \yt) components as peak structures.
(b)  Correlated-pair density on 2D angular difference space $(\eta_\Delta,\phi_\Delta)$. Hadrons are selected with $p_t \approx 0.6$ GeV/c ($y_t \approx 2$). Nevertheless, features expected for dijets are observed: (i) same-side 2D peak representing intrajet correlations and (ii) away-side 1D peak on azimuth representing interjet (back-to-back jet) correlations~\cite{porter2,porter3}. 
 }  
 \end{figure}

Figure~\ref{ppcorr} (left panel) shows $y_t \times y_t$ correlations for 200 GeV approximately NSD \pp\ collisions. The logarithmic interval $y_t \in [1,4.5]$ corresponds to $p_t \in [0.15,6]$ GeV/c. The two peak features correspond to TCM soft and hard components. The 2D hard component with mode near \yt\ = 2.7 ($\approx 1$ GeV/c) corresponds quantitatively to the 1D SP spectrum hard component modeled by $\hat H_0(y_t)$ in Ref.~\cite{ppprd}. The soft component (US pairs only) is consistent with longitudinal fragmentation (dissociation) of projectile nucleons manifesting local charge conservation.

Figure~\ref{ppcorr} (right panel) shows 2D angular correlations on difference variables $(\eta_\Delta,\phi_\Delta)$. The hadron \pt\ values for that plot are constrained to lie near 0.6 GeV/c (just above \yt\ = 2), corresponding to the saddle between soft and hard peaks in the left panel. Although the hadron \pt\ is very low the structures expected for jet angular correlations are still clearly evident: a SS 2D peak at the origin representing intrajet correlations and a 1D peak on azimuth at $\phi_\Delta = \pi$ corresponding to interjet correlations between back-to-back jet pairs. The volume of the SS 2D peak corresponds quantitatively to the hard component of the total hadron yield inferred from $y_t$ spectrum data and to pQCD calculations~\cite{jetspec}. The soft component, a narrow 1D Gaussian on $\eta_\Delta$ including only US charge pairs, is excluded by the $p_t > 0.5$ GeV/c cut~\cite{porter3}. 

Angular correlation systematics have been compared to the QCD Monte Carlo \textsc{pythia}~\cite{pythia}, and general qualitative agreement is observed~\cite{porter2}. 
The correlation measure $\Delta \rho / \sqrt{\rho_{ref}}$ proportional to the number of correlated pairs per final-state hadron~\cite{anomalous} is analogous to ratio $n_h / n_s$ given $n_h \rightarrow$ correlated-pair number. In  the present study we extend  the \pp\ results by measuring systematic variations of 2D angular correlations with parameter $n_{ch}'$.

\begin{figure*}[t]
\includegraphics[width=.3\textwidth]{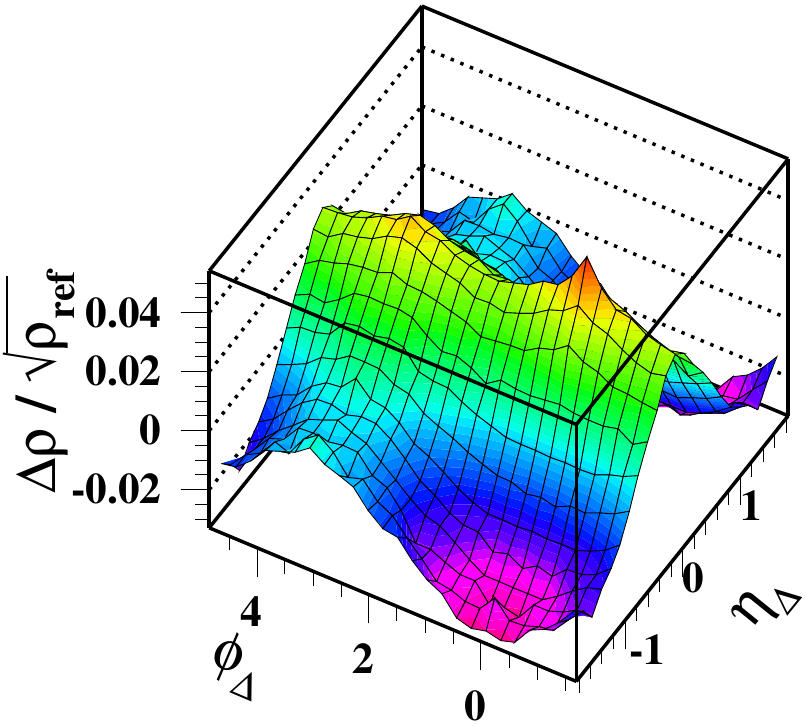}
\put(-120,115) {\bf (a)}
\includegraphics[width=.3\textwidth]{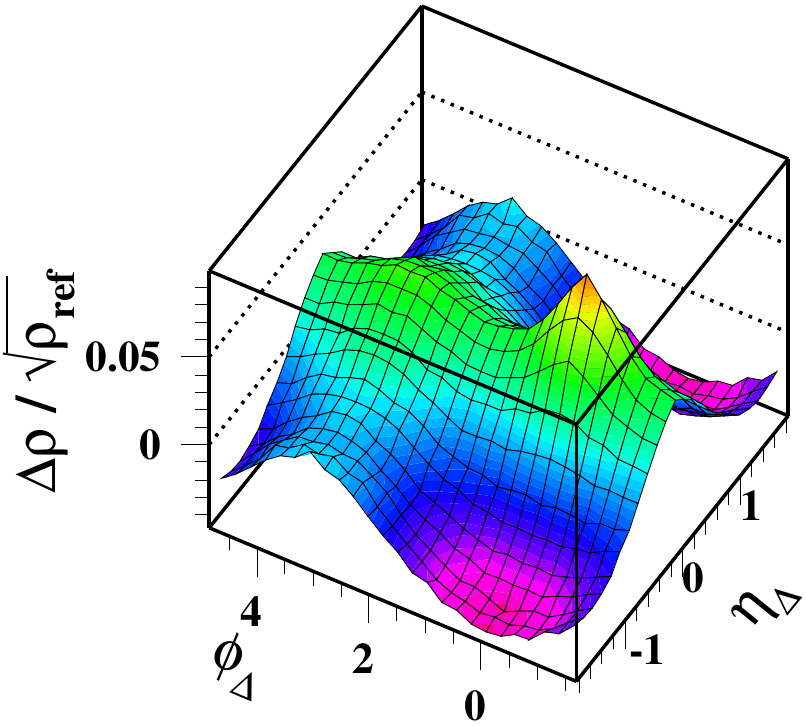}
\put(-120,115) {\bf (b)}
\includegraphics[width=.3\textwidth]{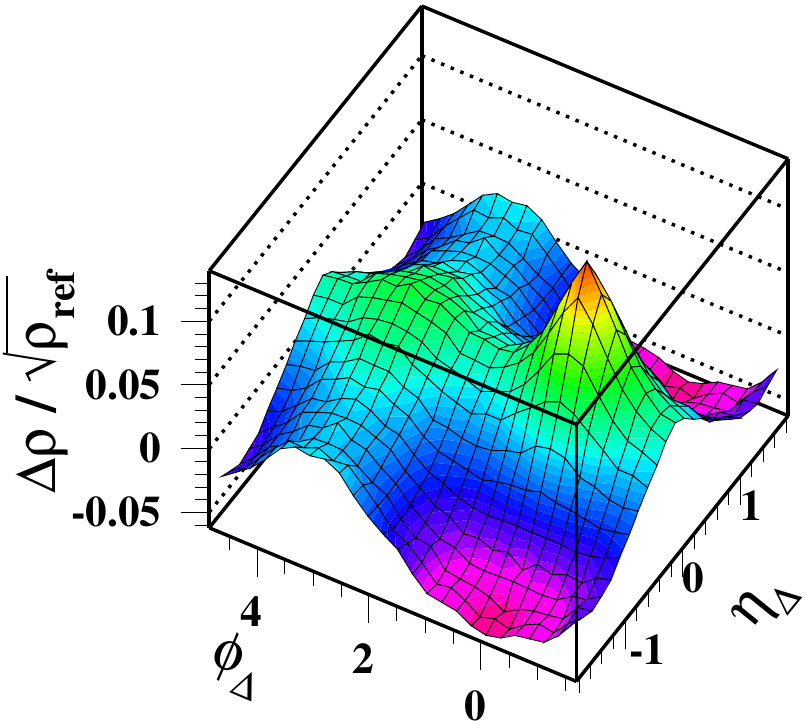}
\put(-120,115) {\bf (c)}\\
\includegraphics[width=.3\textwidth]{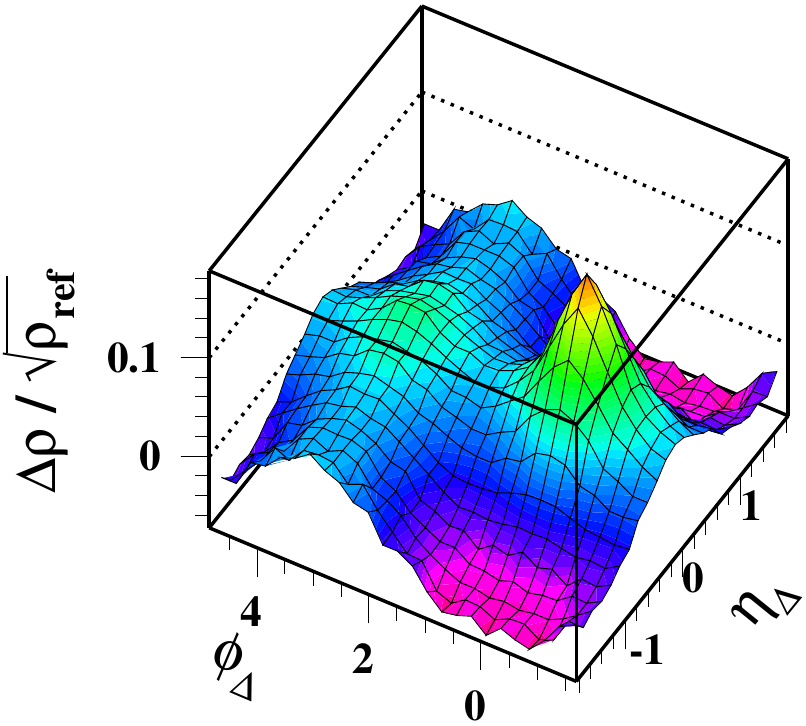}
\put(-120,115) {\bf (d)}
\includegraphics[width=.3\textwidth]{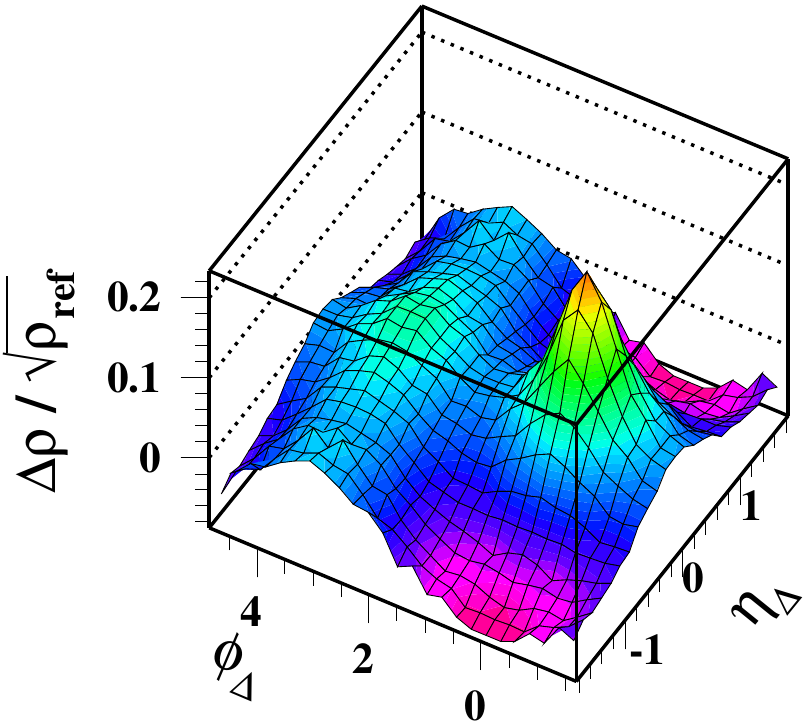}
\put(-120,115) {\bf (e)}
\includegraphics[width=.3\textwidth]{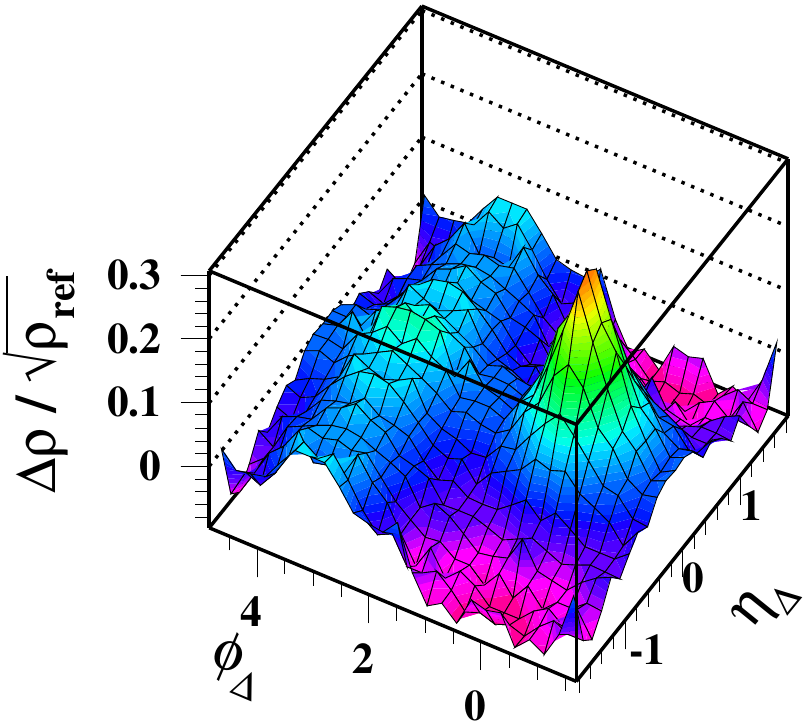}
\put(-120,115) {\bf (f)}
\caption{\label{data} 
(Color online) Evolution of \yt-integral charge-independent 2D angular correlations as $\Delta\rho/\sqrt{\rho_{ref}} \equiv \bar \rho_0 (\bar \rho_{sib} / \bar \rho_{mix} - 1)$ on  $(\eta_{\Delta},\phi_{\Delta})$ with $n_{ch}'$ (see Table~\ref{multclass})  for \pp\ collisions at $\sqrt{s}$ = 200~GeV averaged over acceptance $\Delta \eta = 2$. The pair ratio in the measure definition cancels instrumental effects and $\bar \rho_0$ is corrected for single-particle inefficiencies.
} 
\end{figure*}

 \section{$\bf p$-$\bf p$ 2D Angular Correlations} \label{ppangcorr}

Two-particle angular correlations are obtained with the same basic methods as employed in Refs.~\cite{axialci,ppprd,anomalous,v2ptb}. Data for this analysis were obtained from a MB sample of \pp\ collisions at $\sqrt{s} = 200$~GeV.
 Charged particles were detected with the STAR Time Projection Chamber (TPC). 
 The acceptance was $2\pi$ azimuth, pseudorapidity $|\eta| < 1$, and $p_t > 0.15$~GeV/c. 
The {\em observed} (uncorrected) charge multiplicity within the $\eta$ acceptance is denoted by $n_{ch}'$, whereas the efficiency-corrected and $p_t$-extrapolated {\em true} event multiplicity in the acceptance is denoted by $n_{ch}$ with corrected mean angular density $\bar \rho_0 = n_{ch} / \Delta \eta$ within acceptance $\Delta \eta$. Seven event classes indexed by the {observed} charged-particle multiplicity are defined in Table~\ref{multclass}. The range of corrected particle density $\bar \rho_0(n_{ch}')$ is approximately 2-20 particles per unit $\eta$. This analysis is based on 6 million (M) events, compared to 3M events for the $p_t$-spectrum study in Ref.~\cite{ppprd}.


\begin{table}[h]
  \caption{Multiplicity classes based on the observed (uncorrected) multiplicity $n_{ch}'$ falling within acceptance $|\eta| < 1$ or $\Delta \eta = 2$. The efficiency-corrected density is $\bar \rho_0 = n_{ch} / \Delta \eta$. Event numbers are given in millions (M = $1 \times 10^6$).  TCM parameters include $\alpha = 0.006$ and $\xi = 0.6$.
}
  \label{multclass}
\begin{center}
\begin{tabular}{|c|c|c|c|c|c|c|c|} \hline
 Class $n$ &1 & 2 & 3 & 4 & 5 & 6 & 7 \\ \hline
 $n_{ch}'$  &2-3& 4-6 & 7-9 & 10-12 & 13-17 & 18-24 & 25-50 \\ \hline
$\langle n_{ch}' \rangle $ & 2.52 & 4.87 & 7.81 & 10.8 & 14.3  & 19.6 & 26.8 \\ \hline
 $\bar \rho_0(n_{ch}')$ & 1.76 & 3.41 & 5.47& 7.56 &  10.0 & 13.7 & 18.8 \\ \hline
Events (M) & 2.31 & 2.21 & 0.91 & 0.33  & 0.14 & 0.02 & 0.001  \\ \hline
\end{tabular}
\end{center}
\end{table}






Table~\ref{multclass} shows the multiplicity classes and numbers of events per class from 6M \pp\ events, with charge multiplicity averaged over acceptance $|\eta|< 1$ or $\Delta \eta = 2$. The first row presents the bins defined on observed (uncorrected) multiplicity $n_{ch}'$  within $\Delta \eta$. The second row presents uncorrected event-number-weighted bin means. The third row presents efficiency- and $p_t$-acceptance-corrected mean densities on $\eta$, with correction factor $\approx 1.6$. Those multiplicity classes extend over a 11:1 ratio range compared with 12:1 for the RHIC \pp\ analysis in Ref.~\cite{ppprd} and 7.6:1 for the LHC analysis in Ref.~\cite{cms}.

The approximately 2.25M events in each of the first two multiplicity bins can be compared with 0.12M events for the 10\% most-peripheral centrality bin from a correlation analysis of 200 GeV \auau\ collisions in Ref.~\cite{anomalous}. The 2D histogram per-bin statistical uncertainties for the present study are more than 4.5 times smaller. Histograms for the first five \pp\ bins each have better statistics than a 200 GeV \auau\ histogram from Ref.~\cite{anomalous}.
%


\begin{figure*}[t]
\includegraphics[width=.24\textwidth]{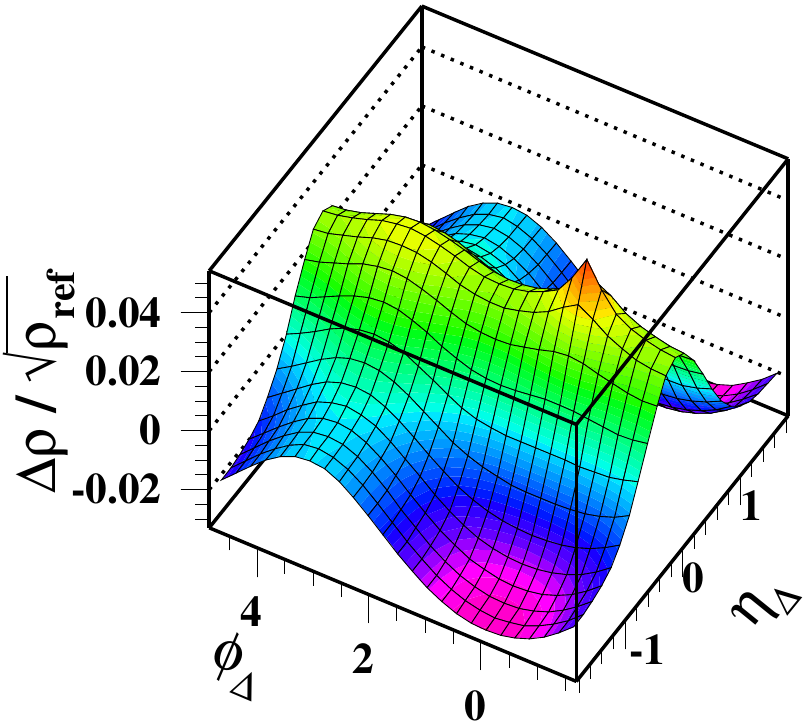}
\put(-100,85) {\bf (a)}
\includegraphics[width=.24\textwidth]{ppcms23-0bx}
\put(-100,85) {\bf (b)}
\includegraphics[width=.24\textwidth]{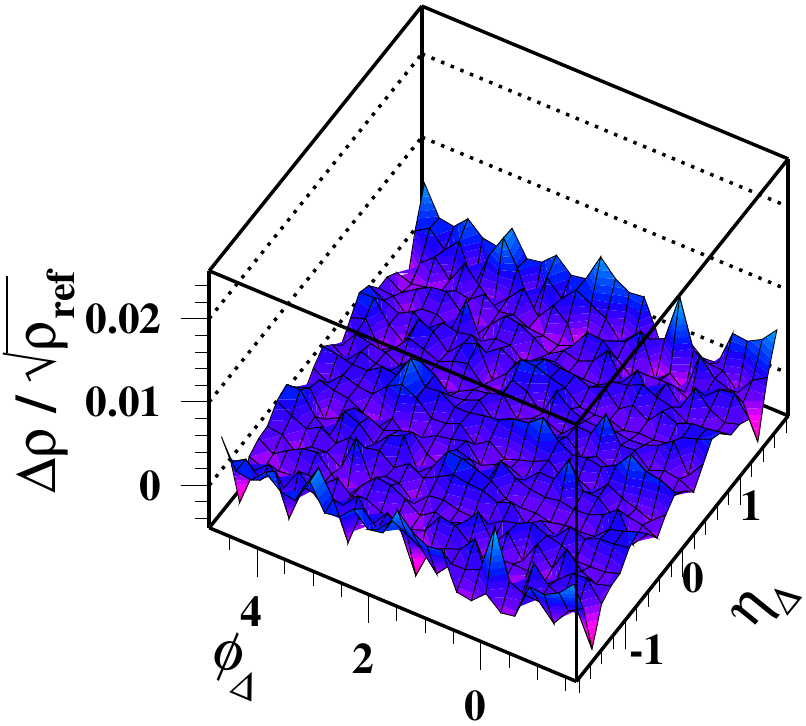}
\put(-100,85) {\bf (c)}
\includegraphics[width=.24\textwidth]{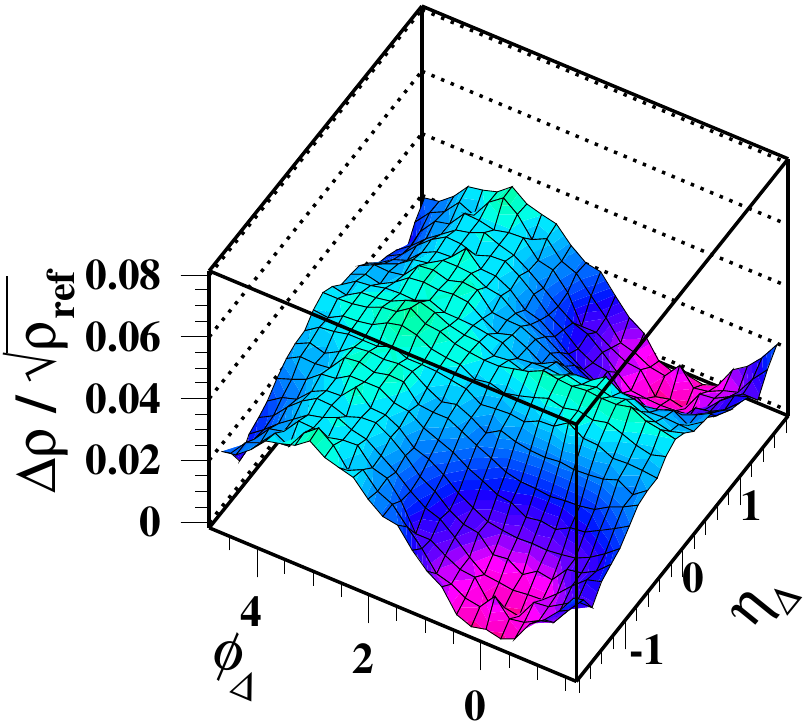}
\put(-100,85) {\bf (d)}\\
\includegraphics[width=.24\textwidth]{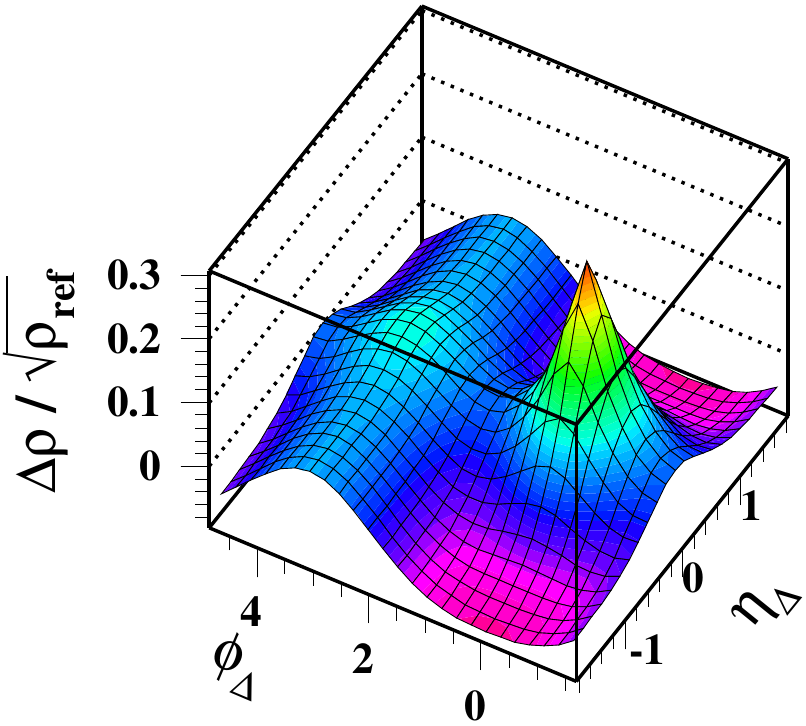}
\put(-100,85) {\bf (e)}
\includegraphics[width=.24\textwidth]{ppcms23-5bx}
\put(-100,85) {\bf (f)}
\includegraphics[width=.24\textwidth]{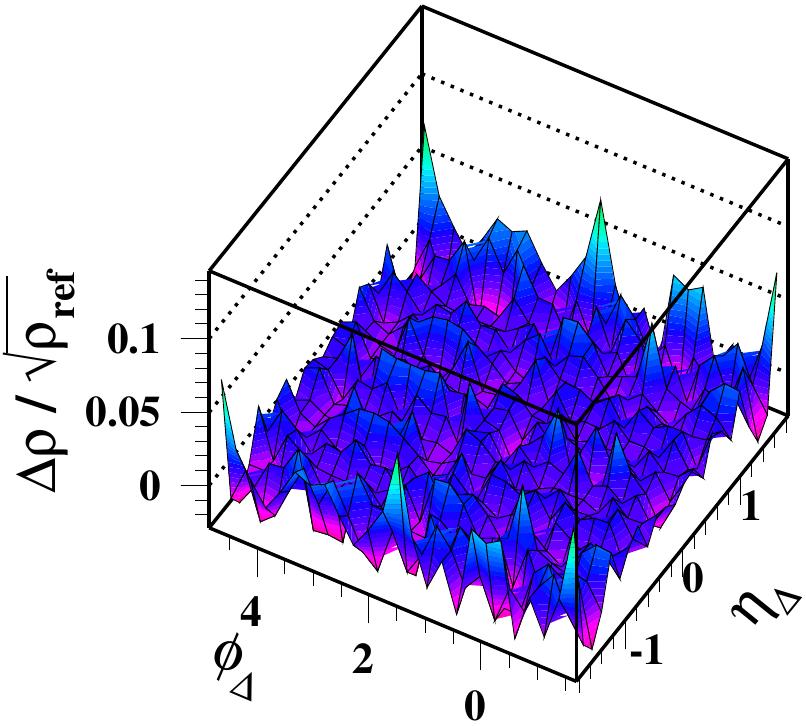}
\put(-100,85) {\bf (g)}
\includegraphics[width=.24\textwidth]{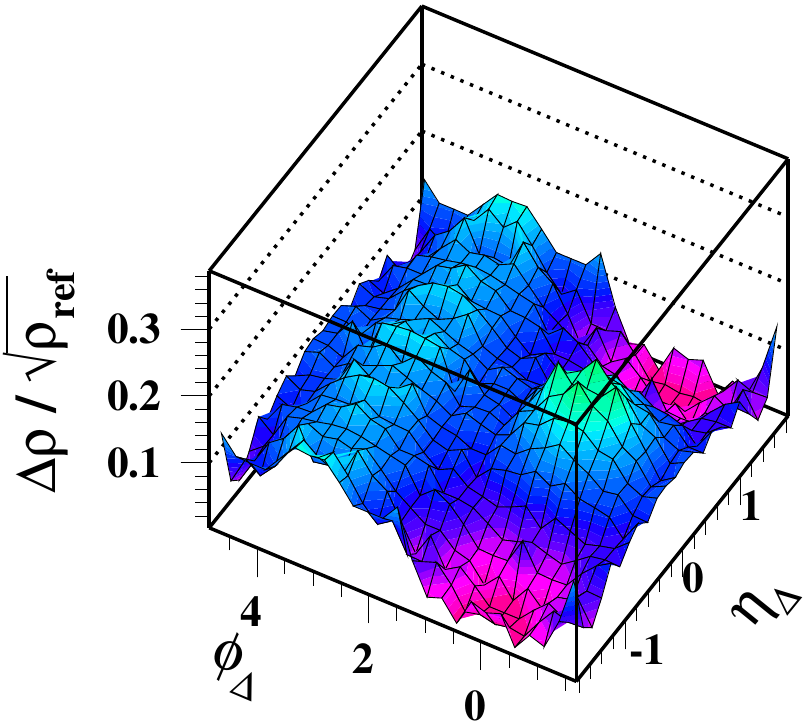}
\put(-100,85) {\bf (h)}
\caption{\label{fits} 
(Color online) Perspective views of charge-independent, \yt-integral 2D angular correlations as $\Delta\rho/\sqrt{\rho_{ref}}$ on  $(\eta_{\Delta},\phi_{\Delta})$  from  \pp\ collisions at $\sqrt{s}$ = 200~GeV for $n=1,~6$ multiplicity classes (upper and lower rows respectively). (a,e) fit model, (b,f) data histogram, (c,g) fit residuals (vertical sensitivity increased two-fold), (d,h) jet + NJ quadrupole contributions obtained by subtracting fitted offset, soft-component and BEC/electron elements of the fit model from the data histograms (see text). 
}  
\end{figure*}

Figure~\ref{data} shows 2D histograms for the first six multiplicity bins. The histograms are corrected for \yt\ and angle-averaged inefficiencies. The correlation measure is $\Delta \rho  / \sqrt{\rho_{ref}} \equiv \bar \rho_0 (\bar \rho_{sib}/\bar \rho_{mix} - 1)$ with prefactor $\bar \rho_0$ corrected~\cite{anomalous}. These are $p_t$-integral or MB angular correlations with no $p_t$ ``trigger'' condition. The 2D histograms are binned $25 \times 25$ on $(\eta_\Delta,\phi_\Delta)$ as for the analyses in Refs.~\cite{axialci,anomalous}.  The pair acceptance on $\eta_\Delta$ (accepted interval on $\eta_\Sigma$) falls linearly from a maximum at the origin to 1/25 that value at the outermost bins. The outermost bins (at $|\eta_\Delta| \approx 2$) are not shown in the figure because statistical fluctuations there distract from the significant correlation structure in other bins.

The prominent correlation features are a 1D peak on $\eta_\Delta$, a SS 2D peak at the origin and an AS 1D peak at $\pi$ on $\phi_\Delta$. The amplitudes of the last two features appear to increase with $n_{ch}$ much more rapidly than that of the first feature. The SS 2D peak structure is actually a superposition of a broader jet-related peak (mainly US pairs) and a narrower composite peak including Bose-Einstein correlations (BEC, LS pairs) and gamma-conversion electrons (US pairs)~\cite{porter2,porter3}. Visual evidence for a significant NJ quadrupole component is also apparent in the figure, as discussed in Sec.~\ref{modelfit}.

\section{2D Model fits} \label{modelfits}

Model fits to 2D angular correlations can provide an accurate quantitative description of 2D histogram data. The fit model resolves several correlation components subsequently interpreted to represent distinct hadron production mechanisms. In this section we apply the  eleven-parameter fit model  from Refs.~\cite{porter2,porter3,axialci,anomalous}.


\subsection{2D model function} \label{2dmodel}

2D angular-correlation histograms are fitted with a {\em six-element model function}, the simplest model that provides a reasonable description of all minimum-bias \pp\ data and \auau\ data for all centralities~\cite{axialci,ppprd,anomalous,v2ptb}. This ``standard'' model was motivated by the simple features apparent in the 2D correlation histograms, not by {\em a priori} physical assumptions~\cite{porter2,porter3}. 
A $\cos(2\,\phi_{\Delta})$ azimuth quadrupole component required to describe Au-Au data is retained in the \pp\ fit model for the present study.

The 2D model function on $(\eta_{\Delta},\phi_{\Delta})$ employed for this analysis includes  (i) a SS 2D Gaussian,  (ii) an $\eta_{\Delta}$-independent AS azimuth dipole $\cos(\phi_\Delta - \pi)$, (iii) an  $\eta_{\Delta}$-independent azimuth quadrupole $\cos(2\, \phi_\Delta)$, (iv) a $\phi_\Delta$-independent 1D Gaussian on $\eta_{\Delta}$, (v) a SS 2D exponential and (vi) a constant offset. 
The 2D fit model is then the sum of the six elements 
%
%
\bea \label{modelfunc}  
\frac{\Delta \rho}{\sqrt{\rho_{\text{ref}}}} 
  & = & A_0 +
A_{2D} \, \exp\left\{- \frac{1}{2} \left[ \left( \frac{\phi_{\Delta}}{ \sigma_{\phi_{\Delta}}} \right)^2  + \left( \frac{\eta_{\Delta}}{ \sigma_{\eta_{\Delta}}} \right)^2 \right] \right\} \nonumber \\ 
&& \hspace{-.3in}+~ A_{\rm D}\,  \{1 +\cos(\phi_\Delta - \pi)\} / 2 + A_{BEC, e\text{-}e}(\eta_\Delta,\phi_\Delta)  \nonumber \\
& & \hspace{-.3in} +~ A_{\rm Q}\, 2\cos(2\, \phi_\Delta)
+A_{\rm soft}\, \exp\left\{-\frac{1}{2} \left( \frac{\eta_{\Delta}}{ \sigma_{\rm soft}} \right)^2 \right\},
\eea
the same eleven-parameter model used to describe MB \auau\ angular correlations in Ref.~\cite{anomalous}. The symmetrized $25\times 25$-bin data histograms include more than 150 degrees of freedom and strongly constrain the model parameters. Based on measured parameter trends and comparisons with QCD theory elements (i) and (ii) together have been attributed to dijet production~\cite{anomalous}, (iii) is conventionally identified with elliptic flow in \aa\ collisions~\cite{2004,gluequad,nohydro}), (iv) is attributed to projectile-nucleon dissociation and element (v) models BEC and conversion-electron pairs. 

Equation~(\ref{modelfunc}) differs from the model in Refs.~\cite{anomalous,pptheory} where the dipole and quadrupole terms are expressed as $A_{\rm D}\, \cos(\phi_\Delta - \pi)$ and $A_{\rm Q}\, \cos(2\, \phi_\Delta)$.  The form of the dipole term in Eq.~(\ref{modelfunc}) is the limiting case (with increasing peak width) of a periodic AS 1D peak array~\cite{tzyam}. Note that in Fig.~\ref{data} the SS 2D peak is fully resolved on $\eta_\Delta$ in all cases. The $m = 1$, 2 multipoles (AS dipole, NJ quadrupole) are then the only other significant structures on azimuth $\phi_\Delta$.



 \subsection{Model-fit results} \label{modelfit}

Fig.~\ref{fits} shows examples of fit decomposition and residuals using data from the first (upper row) and sixth (lower row) multiplicity bins.  The two rows show (left to right) model fit, data, residuals (data$-$model) and jet-related + NJ quadrupole structure. The last are obtained by subtracting fitted model elements soft (iv), BE/electrons (v) and offset (vi) from the data histograms leaving jet-related structure (i), (ii) and NJ quadrupole (iii). Residuals (c) and (g) (those plots are a factor 2 more sensitive) are comparable in magnitude to statistical errors and negligible compared to the amplitudes of the principal correlation features. The lack of significant nonrandom structure in the residuals suggests that the standard fit model exhausts all information in the data.  For comparison purposes panels (d) and (h) are plotted with the same vertical intervals as data histograms (b) and (f), but the fitted offsets (vi) have been subtracted. 

 \begin{figure}[h]
  \includegraphics[width=1.65in,height=1.6in]{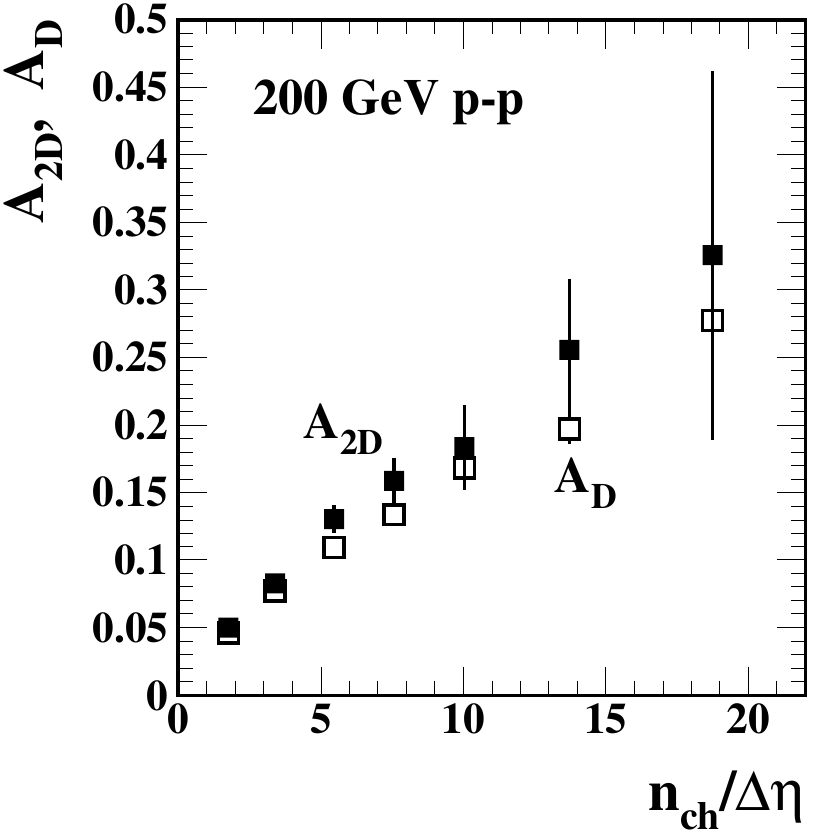}
  \includegraphics[width=1.65in,height=1.6in]{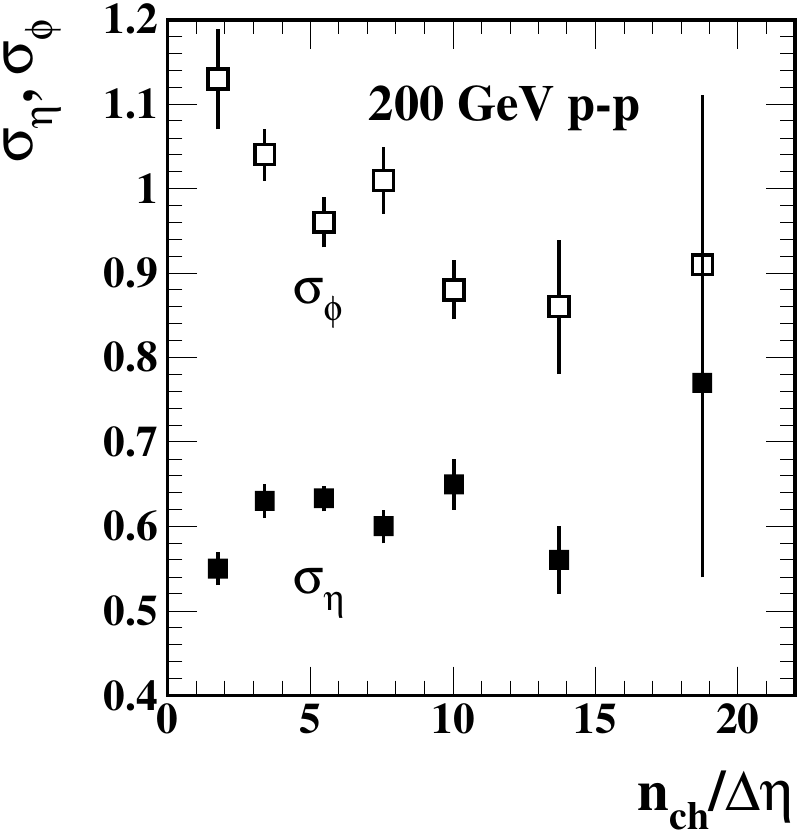}
\caption{\label{ss2dpar}
Left: SS 2D peak and AS 1D dipole amplitudes $A_{2D}$ and $A_D$ vs corrected charge density $\bar \rho_0 = n_{ch} / \Delta \eta$. 
Right: SS 2D peak $\eta$ and $\phi$ widths vs charge density.  
 }  
 \end{figure}

\begin{figure}[h]
  \includegraphics[width=1.65in,height=1.6in]{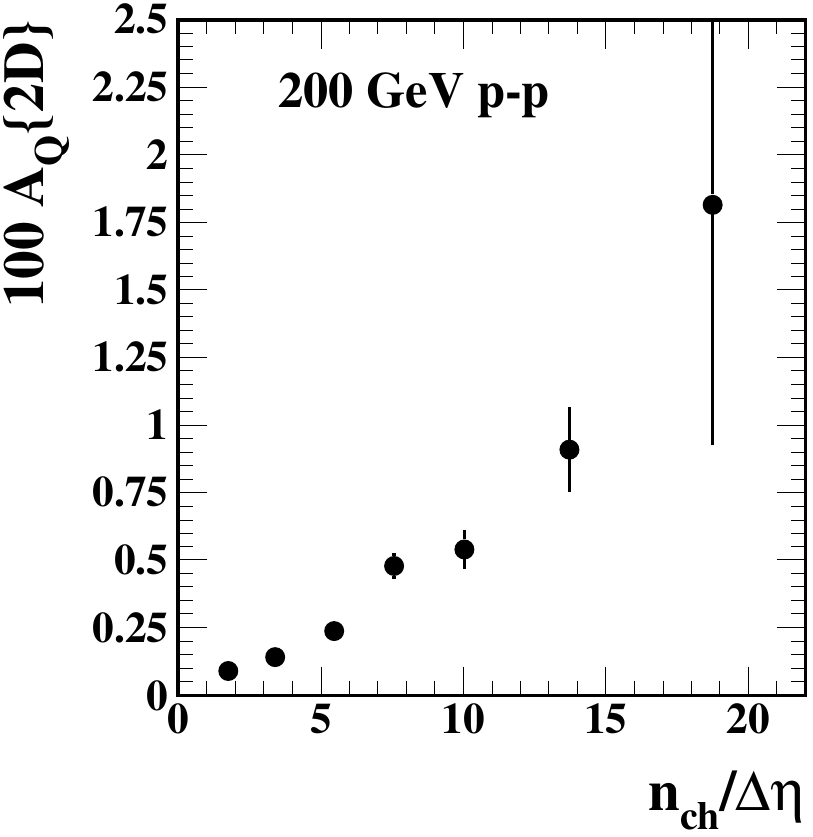}
  \includegraphics[width=1.65in,height=1.6in]{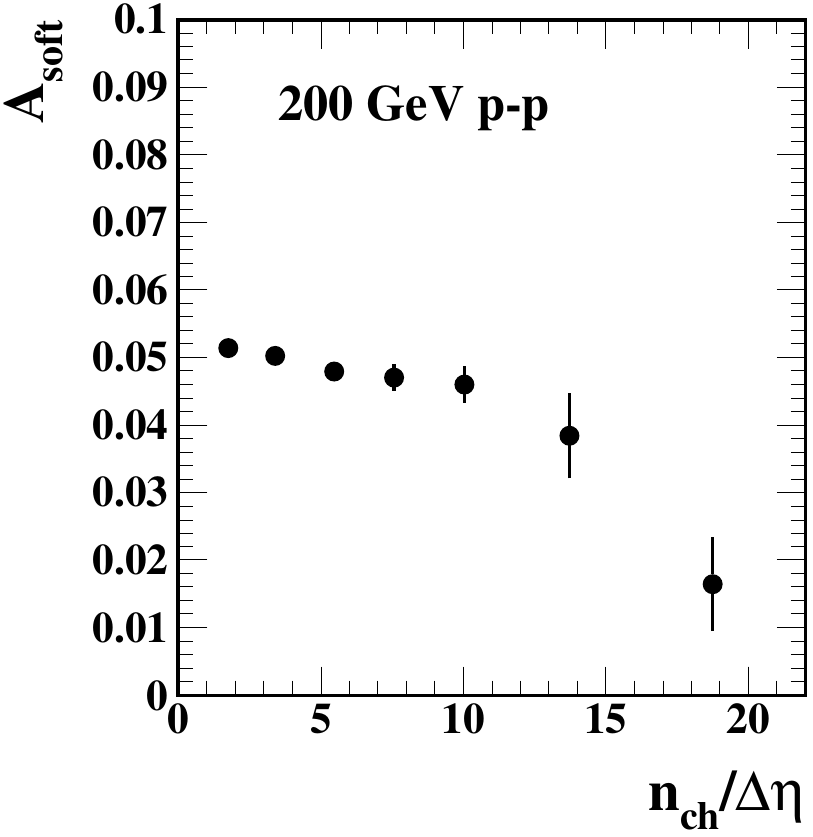}
\caption{\label{quadamp1}
Left:  NJ quadrupole amplitude $ A_Q\{2D\} \equiv \bar \rho_0\, v_2^2\{2D\}$ vs corrected charge density. 
Right: Soft-component 1D Gaussian (on $\eta_\Delta$) amplitude $A_{\rm soft}$ vs charge density.
 }  
 \end{figure}

Figures~\ref{ss2dpar} and \ref{quadamp1} show fit-parameter trends vs corrected multiplicity density $n_{ch} / \Delta \eta = \bar \rho_0$. Best-fit descriptions of the data histograms were obtained with a $\chi^2$ minimization procedure. The general trends with increasing $n_{ch}$ are:  
(a) {\em per-particle} SS 2D and AS 1D peak amplitudes increase approximately linearly with $n_{ch}$, 
(b) the NJ quadrupole amplitude increases approximately as $n_{ch}^2$ and
(c) the soft-component amplitude remains constant.
The SS-peak per-particle amplitude increases nearly ten-fold with similar increase of $n_{ch}$, consistent with the spectrum hard-component scaling trend reported in Ref.~\cite{ppprd}.

The SS 2D peak remains {\em strongly elongated on azimuth} even for larger multiplicities (aspect ratio drops from 2:1 to 1.5:1) in contrast to strong elongation on $\eta$ in more-central \auau\ collisions~\cite{axialci,anomalous}. The width of the soft-component 1D peak on $\eta_\Delta$ is $\sigma_\text{soft} = 0.47 \pm 0.02$, comparable to the $\eta_\Delta$ width of the SS 2D peak $\sigma_{\eta_\Delta} \approx 0.6$. 
The correlated {\em pair number} from the BEC part of the SS peak is expected to scale quadratically with $n_{ch}$. The per-particle BEC amplitude should then increase approximately linearly with $n_{ch}$, similar to observed jet-related correlations from \pp\ collisions. That expectation is consistent with the data in Fig.~\ref{data}.

Visible manifestations of the rapidly-increasing NJ quadrupole amplitude are apparent in Fig.~\ref{fits}. The SS azimuth {\em curvature} within $|\eta_\Delta| > 1$ at $\phi_\Delta = 0$ is substantially positive in (b) or (d) but becomes negligible in (f) or (h), whereas the negative AS curvature near  $\phi_\Delta = \pi$ approximately doubles with \nch. The combined trends are expected with the presence of an increasing azimuth quadrupole component. The relation between NJ quadrupole, AS dipole and resulting SS curvature is discussed further in Sec.~\ref{ridgecms} in connection with the SS ``ridge'' observed in LHC \pp\ angular correlations for certain cut conditions, as first reported in Ref.~\cite{cms}.

To summarize, the per-particle phenomenology of 2D angular correlations vs $n_{ch}$ for 200 GeV \pp\ collisions as in Fig.~\ref{data} includes three main components (soft, hard, NJ quadrupole) exhibiting simple trends with \nch:
(a) The soft-component amplitude remains approximately constant,
(b) the hard-component (jet-related) amplitudes increase linearly, and
(c) the NJ quadrupole amplitude increases quadratically and becomes visually apparent for larger \nch.
We first examine the jet-related trends in the context of the TCM and pQCD dijet production, then consider two NJ components and extend the TCM to include the NJ quadrupole component. 



\section{Jet-related correlations} \label{jetcorr1}

We combine pQCD predictions for minimum-bias dijet production in NSD \pp\ collisions from Ref.~\cite{fragevo} with an $n_{ch}$ trend inferred from spectrum data in Ref.~\cite{ppprd} and the relation between spectrum yields and jet-related angular correlations established in Ref.~\cite{jetspec} to predict jet-related angular-correlation trends for 200 GeV \pp\ collisions. We consider components (i) and (ii) of 2D angular correlations as described in Sec.~\ref{2dmodel}.  The plotted quantities are per-hadron correlation measures $A_X$ converted to per-participant-parton measures by the added factor $n_{ch}/n_s = \bar \rho_0(\Delta \eta) / \bar \rho_s(\Delta \eta)$ assuming $\bar \rho_s$ represents participant low-$x$ partons (gluons).


\subsection{Predicting dijet correlations}

In terms of per-particle correlation measure $\Delta \rho / \sqrt{\rho_{ref}}$ the volume of the SS 2D peak averaged over the angular acceptance $(\Delta \eta, \Delta \phi)$ is represented by~\cite{jetspec}
\bea \label{jetcorr1}
J^2 / \rho_0 = \bar \rho_0 j^2(b) \equiv \frac{2\pi \sigma_\eta \sigma_\phi A_{2D}}{2\pi \Delta \eta},
\eea
with $j^2 = 2\epsilon(\Delta \eta)\,  n_j\,  \overline{n^2_{ch,j}(\Delta \eta)} / \overline{n_{ch}(n_{ch} - 1)}$ as an event-wise pair ratio (number of correlated pairs over total number of pairs in some acceptance)~\cite{jetspec}. Equation (\ref{jetcorr1}) is a per-particle measure of jet-correlated pairs from all jets within the angular acceptance. The factor $2\epsilon(\Delta \eta)$ ($\approx 1.3$ for $\Delta \eta = 2$) includes the probability that the recoil partner jet of a dijet also appears within $\Delta \eta$~\cite{jetspec}.


The volume of the SS 2D peak modeled as a 2D Gaussian is represented by ${\rm V}_{\rm SS2D} = 2\pi \sigma_\eta \sigma_\phi A_{2D}$. The SS peak volume includes any $n_{ch}$ dependence of the peak widths and is thus a more direct measure of jet fragment yields than $A_{2D}$. The results above can be combined to obtain
\bea \label{vss}
V_{SS2D}&=& 2 \pi \Delta \eta \bar \rho_0 j^2 = n_{ch} j^2
\\ \nonumber
&=& n_{ch} 2 \epsilon(\Delta \eta) n_j(n_{ch}) \frac{\overline{n_{ch,j}^2}}{n_{ch}^2}.
\eea
The per-participant-parton measure is then
\bea \label{vss}
\frac{n_{ch}}{n_s}V_{SS2D}&=& 2 \epsilon n_j \overline{n_{ch,j}^2} \times 1/n_s
\\ \nonumber &=& 2\epsilon(\Delta \eta) f_{NSD}  \frac{\bar n_{ch,j}^2(\Delta \eta)}{\bar \rho_{s,NSD}^2}G^2(\Delta \eta) \bar \rho_s 
\eea
where $O(1)$ factor $G^2 = \overline{n_{ch,j}^2} / \bar n_{ch,j}^2$ accounts for fluctuations in the jet fragment multiplicity (see App.~C of Ref.~\cite{jetspec}), and we have used the expression for $f(n_{ch}') = n_j(n_{ch}') / \Delta \eta$ in Eq.~(\ref{nj1}) including a value for $f_{NSD}$ consistent with event-wise-reconstructed jet data.
 
\subsection{Data trends for jet-related correlations}

Factor $n_{ch}/n_s$ applied to fitted correlation amplitudes from histograms in Fig.~\ref{data} in the context of Ref.~\cite{pptheory} converts from per-charged-hadron to per-participant-parton measures, analogous to factor $2n_{ch} /N_{part}$ applied to spectra in \aa\ analysis~\cite{hardspec}. We then plot renormalized amplitudes vs $\bar \rho_s$ to test the relation predicted by Eq.~(\ref{vss}).

 \begin{figure}[h]
  \includegraphics[width=1.65in,height=1.6in]{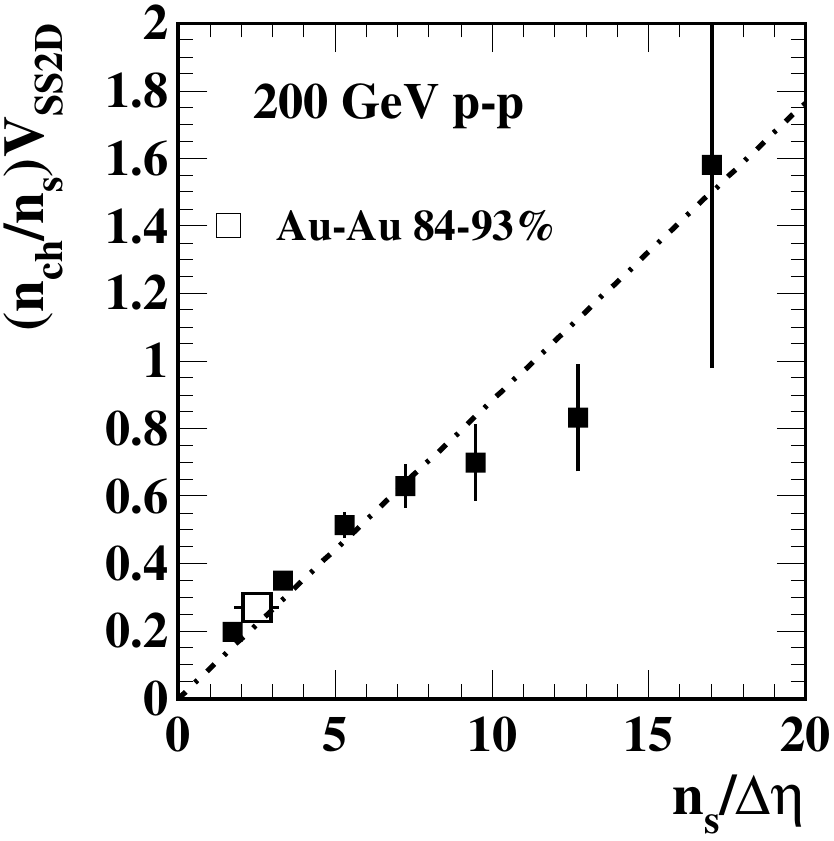}
  \includegraphics[width=1.65in,height=1.63in]{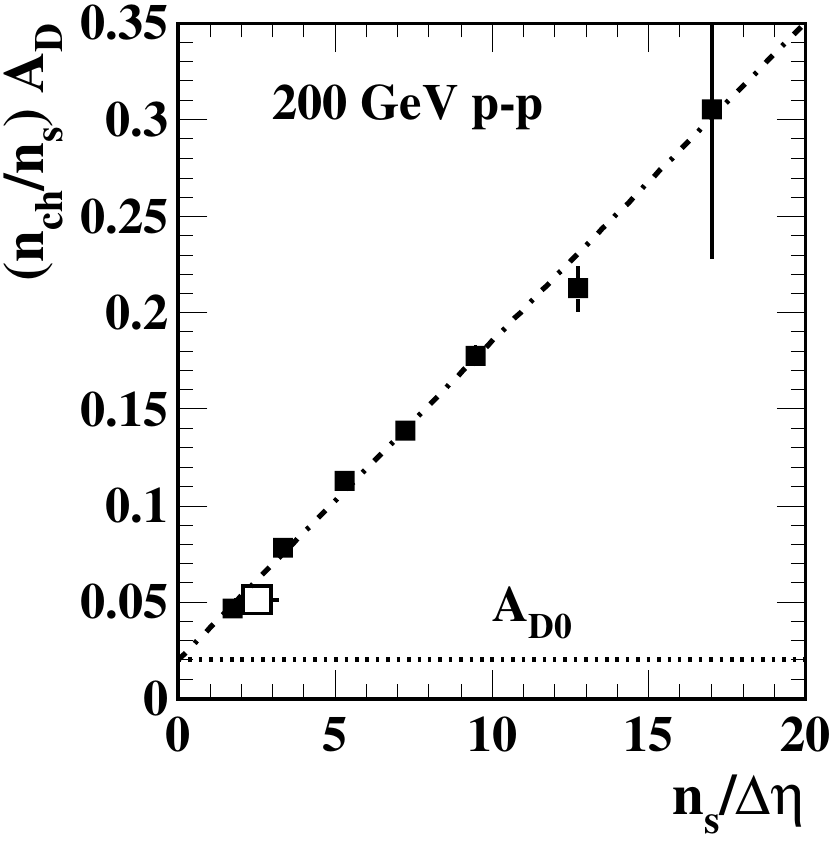}
\caption{\label{ppcent1}
Left: SS peak volume $V_{SS2D}$ rescaled by factor $n_{ch} / n_s$ to per-participant form (for $\Delta \eta = 2$) vs soft-component density $\bar \rho_s \equiv n_s / \Delta \eta$.
Right: Rescaled AS dipole amplitude $A_D$ vs soft-component density. Constant offset $A_{D0}$ is expected for global transverse-momentum conservation. Open boxes represent peripheral \auau\ data equivalent to \nn\ collisions.
}  
 \end{figure}

Figure~\ref{ppcent1} (left panel) shows per-participant jet-related SS 2D peak volume vs soft multiplicity density $n_s/\Delta \eta = \bar \rho_s$. The solid squares are $V_{SS2D}$ derived from SS peak amplitude $A_{2D}$ combined with measured widths $\sigma_{\eta_\Delta}$ and  $\sigma_{\phi_\Delta}$ from Fig.~\ref{ss2dpar}. 
The open square is derived from analysis of 200 GeV \auau\ collisions~\cite{anomalous}. The 84-93\% centrality bin approximates MB  \nn\ collisions. 

The dash-dotted line represents Eq.~(\ref{vss}) (second line) with mean fragment multiplicity $\bar n_{ch,j} = 3.3$ and fluctuation parameter $G^2 = 1.5$. The fragment multiplicity is scaled up from value 2.2 inferred for $\Delta \eta = 1$ in Ref.~\cite{ppprd} based on simulations that indicate a 50\% jet fragment detection efficiency for $\Delta \eta = 1$ and 75\% efficiency for $\Delta \eta = 2$ (acceptance-edge losses). The fluctuation parameter value is Poisson value 1.3 for $\bar n_{ch,j} = 3.3$ plus an estimated additional contribution from non-Poisson fluctuations (see Ref.~\cite{jetspec}, App.~A). The other factors are $f_{NSD} = 0.027$ from Sec.~\ref{ppdijet} and $2\epsilon = 1.25$ for $\Delta \eta = 2$.

Figure~\ref{ppcent1} (right panel) shows per-participant amplitudes for the AS 1D peak (solid squares) interpreted to represent back-to-back jet pairs. Since the AS 1D peak has a fixed geometry and represents a fixed fraction of the dijet number we expect $A_D \propto V_{SS2D}$ modulo small offset $A_{D0}$ representing global transverse-momentum conservation that is expected to be independent of $\eta$ acceptance and/or charged-particle number~\cite{anomalous}. The dashed line is Eq.~(\ref{vss}) times factor 1/5 adjusted to accommodate the $A_D$ data, confirming that $A_D \propto V_{SS2D}$ within data uncertainties modulo fixed offset $A_{D0} \approx 0.02$.


\section{Nonjet correlations} \label{njcorr}

The nonjet components of 2D angular correlations correspond to three model elements described in Sec.~\ref{2dmodel}: (iii) the NJ azimuth quadrupole, (iv) a 1D peak on $\eta_\Delta$ attributed to projectile nucleon dissociation and (v) a contribution to the SS 2D peak from  BEC and conversion electron pairs. The last is not relevant to this study except as a possible source of systematic error. The first two are considered in this section. As above, the plotted quantities are per-hadron correlation measures $A_X$ converted to per-participant-parton measures by added factor $n_{ch}/n_s = \bar \rho_0(\Delta \eta) / \bar \rho_s(\Delta \eta)$.

\subsection{Predicting the NJ azimuth quadrupole}

For \auau\ collisions we observe the following centrality trend for the NJ quadrupole (number of correlated pairs) valid over a large energy interval above 13 GeV~\cite{davidhq}
\bea \label{aaquad}
V_2^2(b) = \bar \rho_0(b) A_Q(b) &\propto& N_{part}(b) N_{bin}(b) \epsilon_{opt}^2(b),
\eea
where for \aa\ collisions $N_{bin} \propto N_{part}^{4/3}$ is a manifestation of the eikonal approximation within the Glauber model.

For \pp\ collisions we find that, relative to the soft hadron density $\bar \rho_s$, dijet production scales $\propto \bar \rho_s^2$ as described in Sec.~\ref{ppdijet}. We then argue by analogy with \aa\ systematics that for \pp\ collisions $N_{part} \sim \bar \rho_s$ and $\text{dijets} \propto N_{bin}\propto N_{part}^2 \sim \bar \rho_s^2$ (the eikonal approximation is not valid). The corresponding form of Eq.~(\ref{aaquad}) for the NJ quadrupole in \pp\ collisions should then be
\bea
V_2^2(n_{ch}') = \bar \rho_0(n_{ch}') A_Q(n_{ch}') &\propto& N_{part}^3 \langle \epsilon^2 \rangle,
\eea 
since there is apparently no systematic dependence on  impact parameter $b$ and therefore no $b$-dependent $\epsilon(b)$.
With $N_{part} \sim \bar \rho_s$ we then obtain
\bea
V_2^2(n_{ch}') / \bar \rho_s = (\bar \rho_0/\bar \rho_s) A_Q(n_{ch}') &\propto& \bar \rho_s^2
\eea
as a {\em predicted} trend for the NJ quadrupole in \pp\ collisions based on measured \auau\ quadrupole systematics and the assumption that {\em the NJ azimuth quadrupole is a universal feature of all high-energy nuclear collisions}.

\subsection{Data trends for nonjet correlations}

\begin{figure}[h]
  \includegraphics[width=1.65in,height=1.6in]{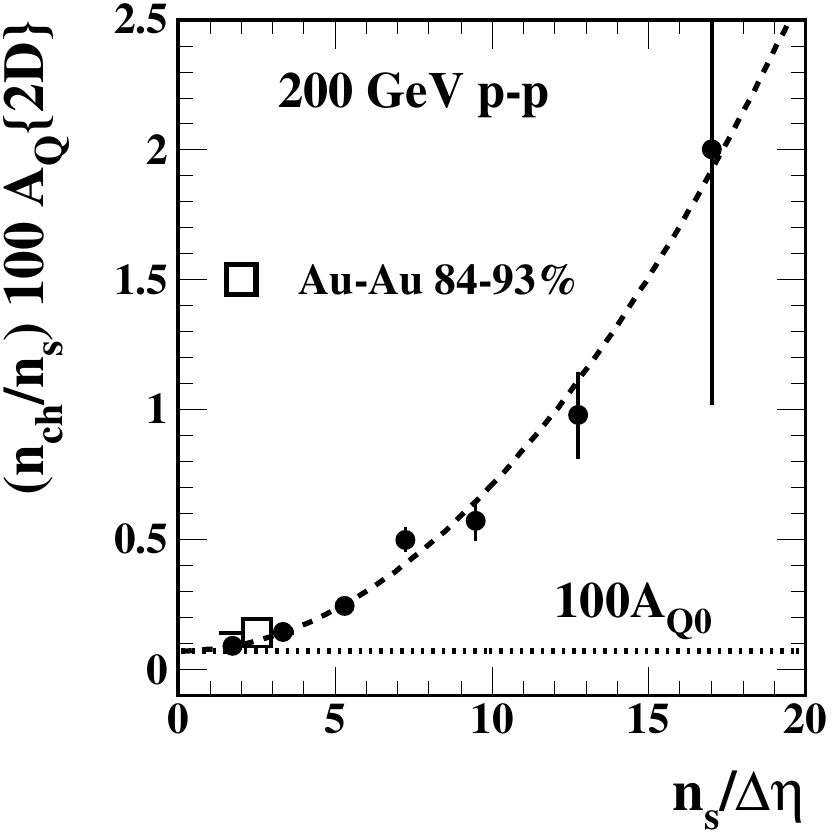}
  \includegraphics[width=1.65in,height=1.6in]{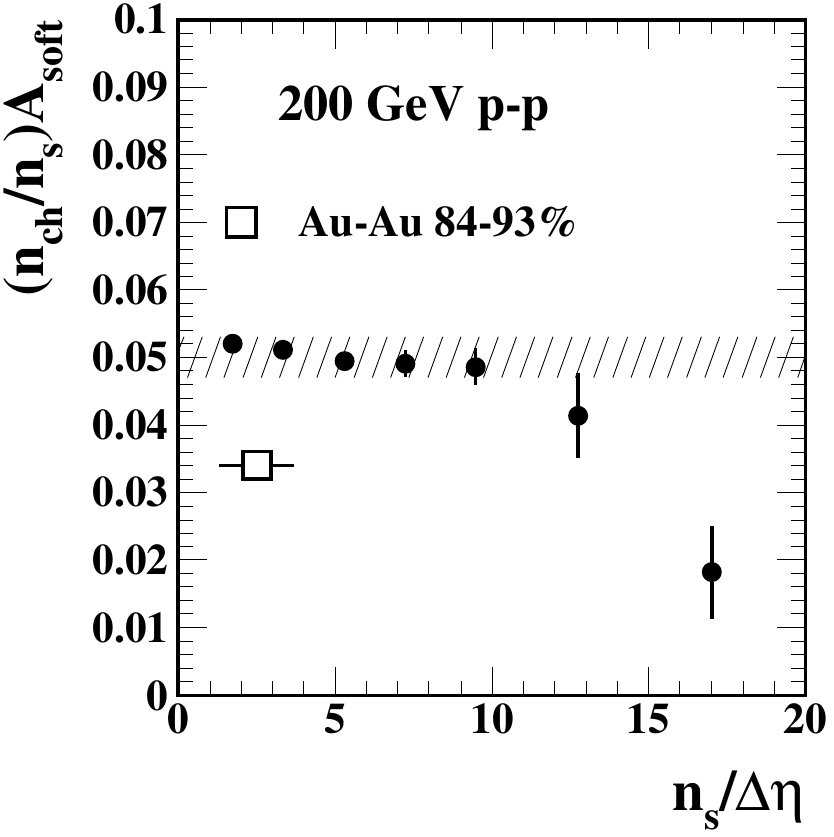}
\caption{\label{soft}
Left: Rescaled NJ quadrupole amplitude $100 A_Q$ vs soft-component charge density. Offset $100 A_{Q0}$ may be an aspect of global transverse-momentum conservation.
Right: Rescaled soft-component amplitude $A_{\rm soft}$ vs soft-component charge density.  The open boxes represent peripheral \auau\ data equivalent to MB \nn\ collisions.
}  
 \end{figure}

Figure~\ref{soft} (left panel) shows fitted NJ quadrupole amplitude $A_Q\{2D\} = \bar \rho_0 v_2^2\{2D\}$ rescaled to a per-participant-parton quantity by factor $\bar \rho_0 / \bar \rho_s = n_{ch} / n_s$. The dashed curve is \bea
(\bar \rho_0 / \bar \rho_s) 100 A_Q\{\text{2D}\} &=& (0.075\bar  \rho_s)^2 + 0.07
\eea indicating that the data (modulo small fixed offset $A_{Q0}$) are consistent with a quadratic dependence on $\bar \rho_s$ in contrast to the linear dependence for jet production in Fig.~\ref{ppcent1}.
The open square is an extrapolation from 200 GeV \auau\ data based on the empirical relation $A_Q\{2D\} = 0.0045 N_{bin} \epsilon_{opt}^2$~\cite{davidhq}. For the peripheral \auau\ limit $\epsilon_{opt} \approx 0.4$ and $N_{bin} = 1$ for \nn\ $\approx$ NSD \pp\ collisions give $100 A_Q\{2D\} \approx 0.07$, in quantitative agreement with \pp\ data (relative to empirical offset $100A_{Q0} =$ 0.07). The offsets $A_{D0}$ and $A_{Q0}$ for dipole and quadrupole components are consistent with global transverse-momentum conservation~\cite{anomalous}. The per-participant trend in Fig.~\ref{soft} (left panel) increases 100-fold over the measured \nch\ interval, and the same quadratic trend on $\bar \rho_s$ continues down to zero hadron density inconsistent with a collective (flow) phenomenon resulting from particle rescattering.

Figure~\ref{soft} (right panel) shows the per-participant soft-component amplitude $(\bar \rho_0 / \bar \rho_s)A_{\rm soft}$. 
Within uncertainties the per-participant amplitude is constant over most of the plotted interval as expected for  correlated-pair number from projectile dissociation $\propto \bar \rho_s$. That trend is consistent with models in which a charged hadron from projectile dissociation may be correlated with one other nearest-neighbor charged hadron, per string fragmentation models and local parton-hadron duality~\cite{lphd}: Nearest-neighbor hadrons are correlated as charge-neutral pairs satisfying local charge and momentum conservation.

The trends of the principal \pp\ correlation components can be summarized as follows (as correlated pairs vs participants): the soft component scales as $\bar \rho_s \propto N_{part}$, the hard (dijet) component scales as $f \propto N_{bin} \sim N_{part}^2$, and the NJ quadrupole scales as $V_2^2 \propto N_{part} N_{bin} \sim N_{part}^3$. In other analysis it was determined that the SS 2D jet peak represented by $A_{2D}$ is comprised mainly of US pairs, whereas the NJ quadrupole in \auau\ collisions is CI (LS = US)~\cite{porter2,porter3}. And the NJ quadrupole and MB dijets have different \nch\ trends. These measurements confirm a revised {\em three}-component model for \pp\ angular correlations in which the NJ quadrupole plays a unique role.

\section{TCM for  $\bf \eta$-density distributions} \label{etadensity}

Section~\ref{pptcm} describes the TCM for \pp\ \yt\ spectra in the form of Eq.~(\ref{ppspec}), a marginal projection onto \yt\ of the 2D hadron density on $(y_t,\eta)$. In this section we consider the complementary marginal projection onto $\eta$ and derive a TCM for the $\eta$ density in terms of averages over a detector acceptance $\Delta \eta$.  Given \yt-spectrum results we expect a correspondence between  the TCM hard component on $\eta$ and the distribution of low-$x$ gluons within the proton.

\subsection{Relation to $\bf y_t$ spectra and TCM} \label{etayt}

Hadron production integrated over azimuth can be represented by SP joint density $ \rho_0(y_t,\eta)$. Given results from the spectrum analysis in Ref.~\cite{ppprd} we assume as a basic TCM decomposition the first line of
\bea \label{eq15}
 \rho_0(y_t,\eta;n_{ch}') &=& S(y_t,\eta;n_{ch}') + H(y_t,\eta;n_{ch}')
\\ \nonumber
&\approx&   \rho_{s0}(n_{ch}') S_0(\eta)  \hat S_0(y_t)
\\ \nonumber
&+&  \rho_{h0}(n_{ch}') H_0(\eta) \hat H_0(y_t).
\eea 
The second line of Eq.~(\ref{eq15}) invokes factorization of both soft and hard components as in Eq.~(\ref{ppspec}). We assume that soft component $S(y_t,\eta;n_{ch}')$ is factorizable within some limited acceptance $\Delta \eta$ consistent with projectile dissociation modeled by string fragmentation. Hard component $H(y_t,\eta;n_{ch}')$ may include significant $\eta$-$y_t$ covariances owing to details of dijet production (strong variation of dijet production with $\eta$). We assume factorization here and consider deviations in Sec.~\ref{etadep}.
Spectrum model functions $\hat S_0(y_t)$ and $\hat H_0(y_t)$ on \yt\ were defined in Sec.~\ref{ppspec1} as in Ref.~\cite{ppprd} and are unit integral (indicated by carets) {\em over the full \yt\ acceptance}.  Model functions $S_0(\eta)$ and $H_0(\eta)$ on $\eta$ are newly defined below, and the $\rho_{x0}$ represent soft- and hard-component hadron densities at $\eta = 0$.


In Ref.~\cite{ppprd} $ \rho_0(y_t,\eta;n_{ch}')$ was averaged over acceptance $\Delta \eta = 1$ to obtain a \yt\ spectrum TCM described by
\bea \label{ytspeceq}
\bar \rho_0(y_t;n_{ch}',\Delta \eta) &\approx& \bar \rho_{s}(n_{ch}',\Delta \eta) \hat S_0(y_t)  
\\ \nonumber
&+& \bar \rho_{h}(n_{ch}',\Delta \eta) \hat H_0(y_t),
\eea
where $\bar \rho_{x}(n_{ch}',\Delta \eta) \equiv n_x / \Delta \eta$ with $x = s$, $h$ and bars indicate averages over $\eta$. That model describes corrected and extrapolated \yt\ spectra. In general, the effects of spectrum low-\yt\ inefficiency and a \yt\ acceptance cutoff must be represented by TCM model functions (compare upper and lower dash-dotted curves in Fig.~\ref{fig1a} -- left). 
Because inferred hard-component model $\hat H_0(y_t)$ is negligible below 0.35 GeV/c we assume that low-\yt\ inefficiencies and cutoff affect only soft components of yields and spectra. We replace unit-integral $\hat S_0(y_t)$ (upper) in Eq.~(\ref{eq15}) with modified soft-component model function $S_0'(y_t)$ (lower) and define soft-component tracking efficiency
\bea \label{eff}
\xi &\equiv& \int_0^\infty dy_t y_t  S_0'(y_t) \leq 1
\eea
implying $\rho'_{s0}(n'_{ch}) = \xi \rho_{s0}(n'_{ch})$.
For the present \yt-integral study we integrate $ \rho'_0(y_t,\eta;n_{ch}')$ over \yt\  to obtain
\bea \label{rho0eta}
\rho_0'(\eta;n_{ch}') &\approx&   \rho_{s0}'(n_{ch}') S_0(\eta)  
+  \rho_{h0}(n_{ch}')  H_0(\eta),
\eea 
where $S_0(\eta)$ and $H_0(\eta)$ are TCM model functions with unit amplitude at $\eta = 0$, and primes indicate the effect of low-\yt\ acceptance limits and inefficiencies. 

Averaging Eq.~(\ref{rho0eta}) over some acceptance $\Delta \eta$ symmetric about $\eta = 0$ gives
\bea \label{etaav}
\bar \rho_0'(n_{ch}',\Delta \eta) &=&  \rho_{s0}'(n_{ch}') \bar S_0(\Delta \eta) 
+ \rho_{h0}(n_{ch}') \bar H_0(\Delta \eta)
 \nonumber \\
&\equiv& \bar  \rho_{s}'(n_{ch}',\Delta \eta)  + \bar \rho_h(n_{ch}',\Delta \eta),
\eea
with mean values $\bar \rho_x(n_{ch}',\Delta \eta) = n_x / \Delta \eta = \rho_{x0} \bar X_0(\Delta \eta)$. The TCM of Eq.~(\ref{rho0eta}) can then be rewritten in the form
\bea \label{rho0eta2}
\rho_0'(\eta;n_{ch}') &\approx&  \bar  \rho_{s}'(n_{ch}',\Delta \eta) \tilde S_0(\eta;\Delta \eta)  
\\ \nonumber
&+& \bar  \rho_{h}(n_{ch}',\Delta \eta)  \tilde H_0(\eta;\Delta \eta)
\eea 
with $\bar \rho_{s}'(n_{ch}',\Delta \eta)$ and $\bar \rho_{h}(n_{ch}',\Delta \eta)$ inferred from $n_{ch}'$ based on an assumed value for parameter $\alpha'$ or $\alpha$ as discussed below.  The $\tilde X_0(\eta;\Delta \eta)$ (denoted by a tilde) have average value 1 over acceptance $\Delta \eta$ by definition.  
Averaging Eq.~(\ref{rho0eta2}) over the $\eta$ acceptance should then be consistent with integration of Eq.~(\ref{ytspeceq}) over the \yt\ acceptance (above the \yt\ cutoff) provided that $\hat S_0(y_t)$ is replaced by $S_0'(y_t)$.

As in Sec.~\ref{ppspec1} we must define a value for parameter $\alpha$ or $\alpha'$ to determine the decomposition of $\bar \rho_0$ into $\bar \rho_h$ and $\bar \rho_s$ or $\bar \rho_s'$. We impose a TCM constraint that with increasing index $n_{ch}'$ no part of $\bar \rho_s$-normalized data distributions should decrease, establishing a lower limit for $\alpha$.  For \yt\ spectra that condition requires a soft component varying approximately as $\bar \rho_s \approx \bar \rho_0 - \alpha \bar \rho_0^2$~\cite{ppprd}.  A self-consistent analysis leads to $\bar \rho_{h}(n_{ch}',\Delta \eta) \approx \alpha(\Delta \eta) \bar \rho_{s}^2(n_{ch}',\Delta \eta) $ for corrected \yt\ spectra, with $\alpha(\Delta \eta = 1) \approx 0.006$ as inferred in Ref.~\cite{ppprd}.
In the present study we define $\alpha'(\Delta \eta) = \bar \rho_{h} / \bar \rho_{s}'^2 = \alpha(\Delta \eta) / \xi^2$ to accommodate a soft-component \yt\ cutoff and low-\yt\ tracking inefficiency. That condition plus $\bar \rho_0' = \bar \rho_s' + \bar \rho_h$ define a quadratic equation from which $n_s'$ or $\bar \rho_{s}' = n_s' / \Delta \eta$ can be inferred for any uncorrected $n_{ch}'$ or $\bar \rho_0'$ given a fixed value for $\alpha'$ or $\alpha$ and efficiency $\xi$.

\subsection{Corrected pseudorapidity densities}

Uncorrected SP $\eta$ densities have the form
\bea
\frac{dn_{ch}'}{d\eta} &=& [1+g(\eta)]\lambda(\eta) \rho_0'(\eta;n_{ch}'),
\eea 
where $g(\eta)$ in the first factor represents a common instrumental distortion antisymmetric about $\eta = 0$ affecting both TCM components in common. The undistorted density $\rho_0'(\eta;n_{ch}')$ is assumed to be symmetric about   $\eta = 0$ given the symmetric \pp\ collision system. Those assumptions permit isolation of $g(\eta)$ for each $ n_{ch}'$ value.

\begin{figure}[h]
  \includegraphics[width=3.3in,height=1.6in]{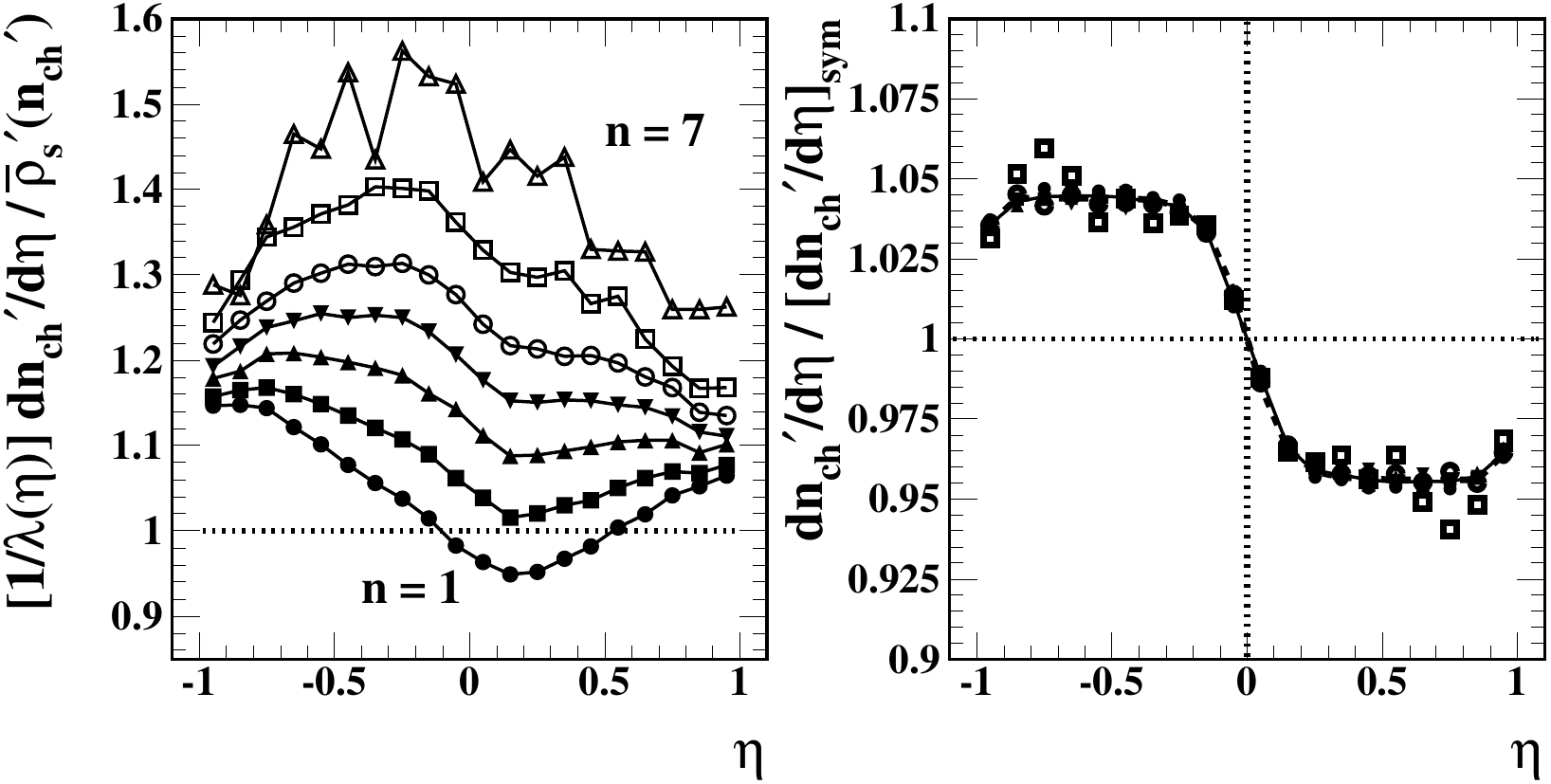}
\caption{\label{dndeta1}
Left: Uncorrected $\eta$ densities within $\Delta \eta = 2$ for seven multiplicity classes (see Table~\ref{multclass}). The curves connecting data points guide the eye.
Right: Ratios of data at left to symmetrized versions revealing a common instrumental asymmetry modeled by the dashed curve.
(data for $n = 7$ are omitted for clarity).
 }  
 \end{figure}

Figure~\ref{dndeta1} (left) shows measured density distributions $dn_{ch}'/d\eta$ normalized by uncorrected soft component  $\bar \rho_s' = n_s' / \Delta \eta$ for several multiplicity classes assuming $\alpha = 0.012$ with efficiency and \yt-acceptance factor $\xi = 0.6$. 
An $\eta$-symmetric  inefficiency $\lambda(\eta) \leq 1$ is observed to deviate from unity only for the outer two bins on each end of the $\eta$ acceptance, with values 0.925 and 0.993 for $|\eta| = 0.95$ and $0.85$ respectively. The plotted data are corrected for  $\lambda(\eta)$
 but a common $\eta$-asymmetric distortion remains. 

Figure~\ref{dndeta1} (right) shows asymmetric data distributions divided by their $\eta$-symmetrized counterparts revealing  the common $\eta$-asymmetric instrumental distortion $g(\eta)$. The dashed model curve  through the points is $1+g(\eta) = 1- 0.044 \tanh(7 \eta)$. The model points at $\eta = \pm 0.95$ are then shifted up and down by 0.008 relative to that trend. The correction is independent of $n_{ch}'$ within statistics. Such an asymmetry is expected if tracking efficiencies in two halves of the TPC differ by a few percent owing to readout-electronics malfunctions. The slope near $\eta = 0$ reflects the primary-vertex distribution on the $z$ axis.

 \begin{figure}[h]
 \includegraphics[width=1.65in,height=1.6in]{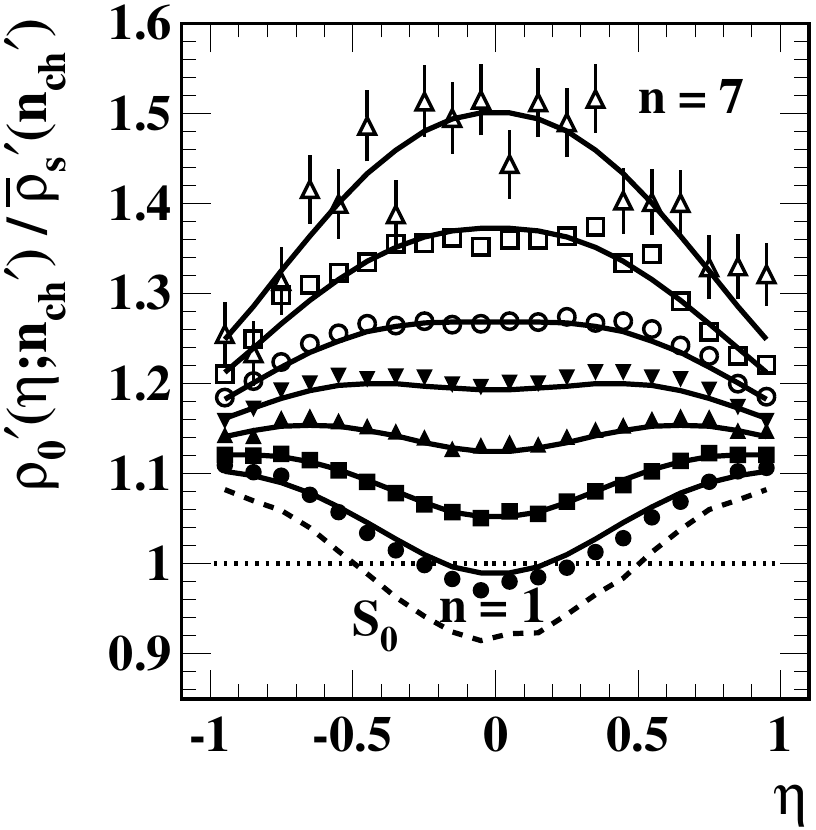}
  \includegraphics[width=1.65in,height=1.6in]{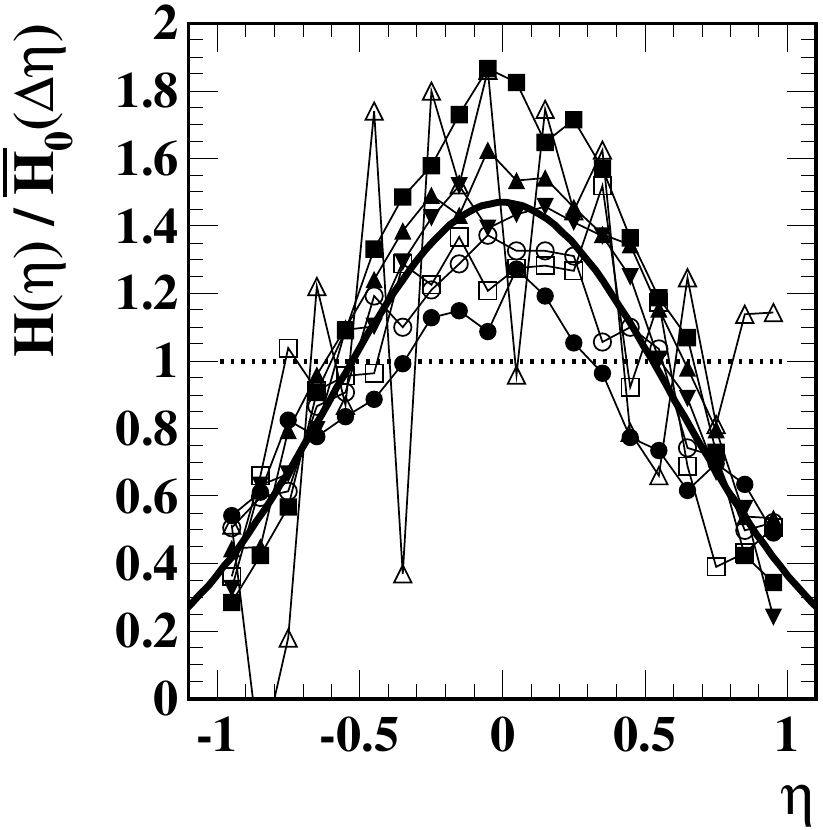}
\caption{\label{dndeta2}
Left:  Corrected $\eta$ densities within $\Delta \eta = 2$ for seven multiplicity classes The dashed curve is normalized soft-component model $\tilde S_0(\eta) \equiv S_0(\eta)/ \bar S_0(\Delta \eta)$ from Eq.~(\ref{s00}). The solid curves are Eq.~(\ref{rho0eta2}) with TCM elements defined in Eqs.~(\ref{s00}) and (\ref{h00}).
Right: Data hard components inferred as in Eq.~(\ref{etahc}). The solid curve is normalized hard-component model $\tilde H_0(\eta)$ as defined in Eq.~(\ref{h00}).
 }  
 \end{figure}

Figure~\ref{dndeta2} (left) shows the asymmetry-corrected normalized densities $\rho_0'(\eta;n_{ch}') /  \bar \rho_{s}'(n_{ch}')$ (points). The solid curves through the data are defined by Eq.~(\ref{rho0eta2}) with soft and hard model functions defined below. The dashed curve is $\tilde S_0(\eta)$ defined in Eq.~(\ref{s00}), the soft-component limit to the ratio $\rho_0'(\eta) / \bar \rho_s'$ as $n'_{ch} \rightarrow 0$. 

\subsection{Inferring the pseudorapidity density TCM}



The TCM soft and hard model functions can be isolated by the following strategy based on Eq.~(\ref{rho0eta2}). Differences between successive ratios $\rho_0'(\eta;n_{ch}') /  \bar \rho_{s}'(n_{ch}')$ (data sets in Fig.~\ref{dndeta2} - left panel) should cancel common term $\tilde S_0(\eta)$ leaving terms approximated by $(\bar \rho_{h,n} / \bar \rho_{s,n}' - \bar \rho_{h,n-1} /   \bar \rho_{s,n-1}') \tilde H_0(\eta)$. The difference data are divided by the factor in parenthesis to obtain estimates for the form of hard-component model $\tilde H_0(\eta)$ which are described by a Gaussian function. $\tilde H_0(\eta)$ is then used to infer $\tilde S_0(\eta)$.

Figure~\ref{dndeta2} (right) shows data difference distributions
\bea \label{etahc}
\frac{H_n(\eta)}{\bar H_0(\Delta \eta)} &\equiv& \frac{\rho_0'(\eta)_{n} /   \bar \rho_{s,n}' - \rho_0'(\eta)_{n-1} /  \bar \rho_{s,n-1}'} {\bar \rho_{h,n} / \bar \rho_{s,n}' - \bar \rho_{h,n-1} /   \bar \rho_{s,n-1}'},
\eea
where index $n \in [1,7]$ represents the seven multiplicity classes. As noted, the difference in the numerator cancels common term $\tilde S_0(\eta;\Delta \eta)$ (defined below). For $n = 1$ we assume $\rho_0'(\eta)_{n-1}/ \bar \rho_{s,n-1}' \rightarrow \tilde S_0(\eta;\Delta \eta)$, and $\bar \rho_{h,n} / \bar \rho_{s,n}' - \bar \rho_{h,n-1} / \bar  \rho_{s,n-1}' \rightarrow \bar \rho_{h,1} / \bar \rho_{s,1}'$ (i.e.\ the extrapolation to $n = 0$ is assumed to be pure soft component). Given the TCM of Eq.~(\ref{rho0eta2}) that expression should represent a common hard-component model in the form $H_0(\eta) / \bar H_0(\Delta \eta)$. The inferred model function for $\Delta \eta = 2$ (bold solid curve) is 
\bea \label{h00}
\tilde H_0(\eta;\Delta \eta) \equiv \frac{H_0(\eta)}{\bar H_0(\Delta \eta)} &=& 1.47 \exp[-(\eta / 0.6)^2/2],
\eea
where  $\sqrt{2\pi \sigma^2} \approx 1.47$ for $\sigma = 0.6$ (the Gaussian tails are truncated) and the normalized function is denoted by a tilde. The hard-component form is, within statistical errors, approximately independent of \nch\ over an interval  implying a 100-fold increase in dijet production, suggesting that most of the hard-component yield  (MB dijet fragments) falls within the acceptance $\Delta \eta = 2$.

 \begin{figure}[h]
  \includegraphics[width=1.65in,height=1.6in]{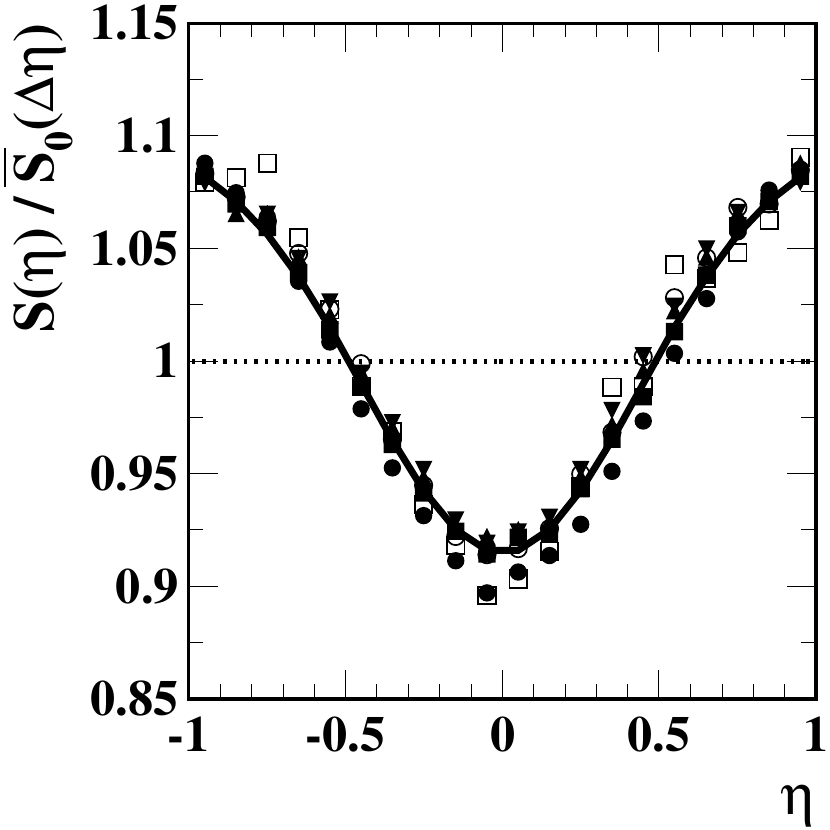}
  \includegraphics[width=1.65in,height=1.6in]{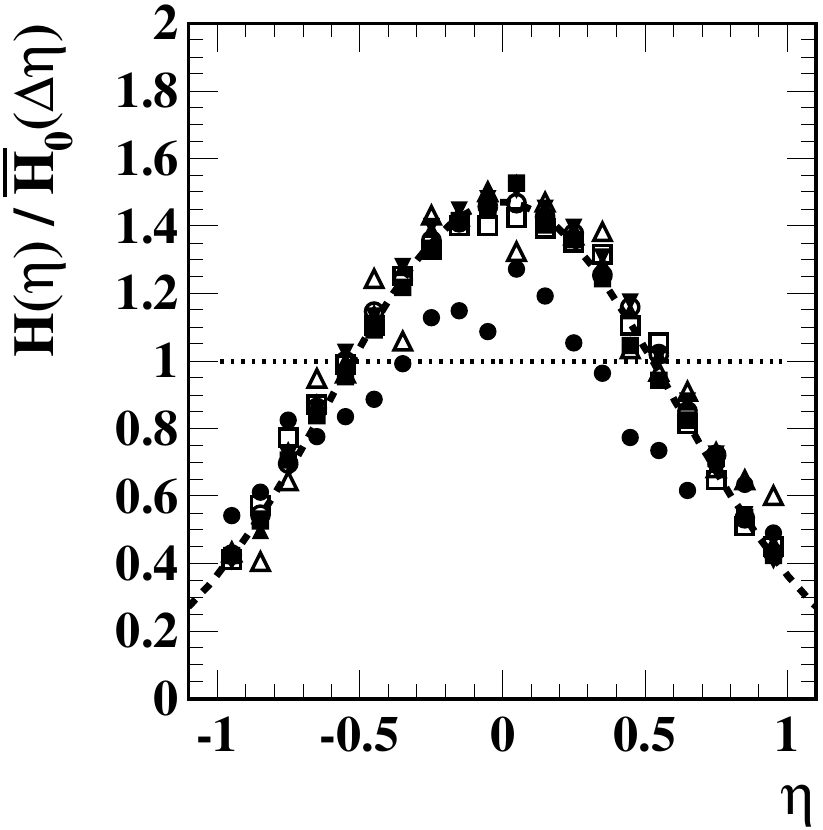}
\caption{\label{etatcm}
Left:  Data soft components inferred as in Eq.~(\ref{ss}). The solid curve is normalized soft-component model $\tilde S_0(\eta)$ as defined in Eq.~(\ref{s00}).
Right:  Data hard components inferred as in Eq.~(\ref{hc01}). The solid curve is normalized hard-component model $\tilde H_0(\eta)$ as defined in Eq.~(\ref{h00}). The data for $n = 1$ (solid dots) are significantly low compared to the common trend.
 }  
 \end{figure}

Figure~\ref{etatcm} (left) shows the soft-component estimator
\bea \label{ss}
\frac{S_n(\eta)}{\bar S_0(\Delta \eta)} &\equiv& \rho_0'(\eta)_{n} /  \bar  \rho_{s,n}' - (\bar \rho_{h,n} / \bar \rho_{s,n}') \tilde H_0(\eta;\Delta \eta),
\eea
with $\tilde H_0(\eta;\Delta \eta)$ as defined in Eq.~(\ref{h00}).  The inferred soft-component model for $\Delta \eta = 2$ (solid curve) is defined by
\bea \label{s00}
\tilde S_0(\eta;\Delta \eta)\hspace{-.02in} \equiv\hspace{-.02in} \frac{S_0(\eta)}{\bar S_0(\Delta \eta)}\hspace{-.05in} &=&\hspace{-.05in} 1.09 - 0.18 \exp[-(\eta / 0.44)^2/2].~~~~~
\eea
The form of the soft component also appears to be invariant over a large \nch\ interval. The small data deviations from the model are consistent with statistical errors. The minimum at $\eta = 0$ is expected given  the Jacobian for $\eta \leftrightarrow y_z$, where an approximately uniform distribution on $y_z$ is expected within a limited $\Delta y_z$ acceptance.

Figure~\ref{etatcm} (right) shows the hard component estimated from data with an alternative method assuming $\tilde S_0(\eta)$
\bea \label{hc01}
\frac{H_n(\eta)}{\bar H_0(\Delta \eta)} &\equiv& \frac{\rho_0'(\eta)_{n} /   \bar \rho_{s,n}' - \tilde S_0(\eta;\Delta \eta)} {\bar \rho_{h,n} / \bar \rho_{s,n}'},
\eea
which substantially reduces statistical noise in  the differences. The dashed curve is the hard-component model $\tilde H_0(\eta;\Delta \eta)$ defined in Eq.~(\ref{h00}) confirming TCM self-consistency. The points for $n=1$ (solid dots) are low compared to the general trend, which may indicate that the hard component for $n=1$ extends sufficiently low on $y_t$ to be significantly reduced by the low-\yt\ tracking inefficiency and acceptance cutoff (see Fig.~\ref{fig1a} -- right).

The hard-component density in the right panel suggests that hadron fragments from MB dijets are strongly peaked near $\eta = 0$, mainly within $\Delta \eta = 2$ consistent with the dominant dijet source being low-$x$ gluons corresponding to small $y_z$ or $\eta$. The functional form on $\eta$ is also consistent with the ``gluon-gluon source'' component described in Ref.~\cite{georg}. The energy dependence $N_{\rm ch}^{gg}\propto \ln^3(s_{NN}/s_0)$ of the integrated gluon-gluon source noted there may arise in \pp\ (\nn) collisions from the $\bar \rho_h \propto \bar \rho_s^2$ trend derived from Ref.~\cite{ppprd} and the present measurements, from the $\bar \rho_s \propto \ln(\sqrt{s} / \text{10 GeV})$ energy trend noted in Ref.~\cite{anomalous}, and from $4\pi$ integration over $\eta$ ($y_z$) that introduces an additional $\ln(\sqrt{s} / \text{10 GeV})$ factor.

\subsection{$\bf y_t$ spectrum $\bf \eta$-acceptance dependence} \label{etadep}

We now consider the effect on the \yt\ spectrum of varying $\Delta \eta$.
As noted below Eq.~(\ref{eq15}) there is reason to expect some $\eta$ dependence for the \yt-spectrum hard component due to the low-$x$ structure of projectile protons (rapid increase of the gluon density with decreasing $x$). Here we present hard components extracted within $\Delta \eta = 1$ and 2 to address that question. We assume that $\hat S_0(y_t)$ does not vary significantly within $\Delta \eta =2$.

 \begin{figure}[h]
  \includegraphics[width=1.65in,height=1.6in]{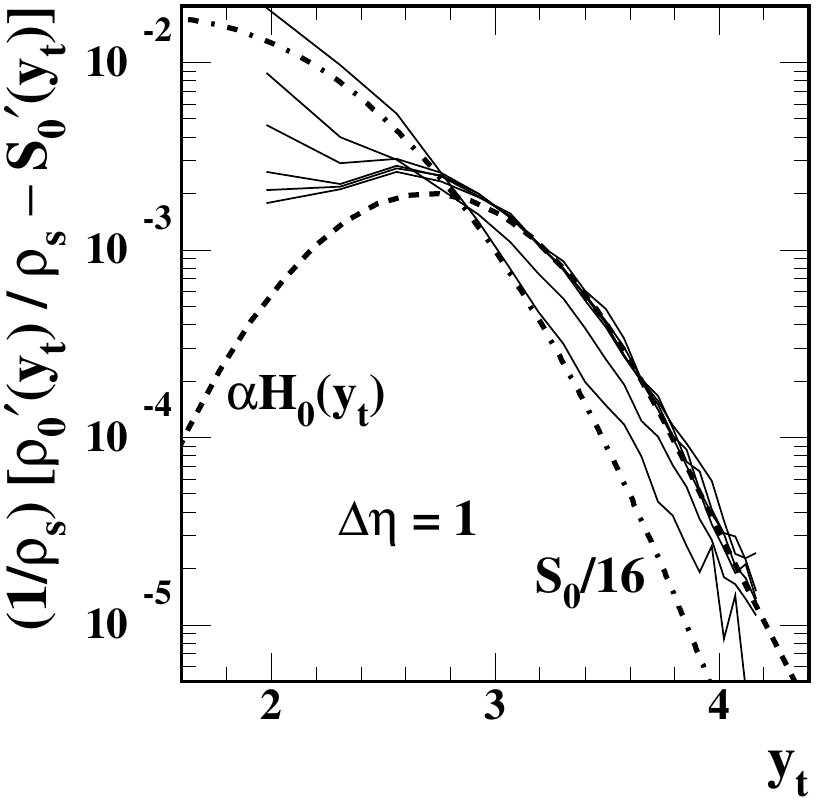}
\includegraphics[width=1.65in,height=1.6in]{ppcms110bfullx}
\caption{\label{etacomp}
Comparison of inferred spectrum hard components for $\Delta \eta = 1$ (left) and $\Delta \eta = 2$  (right) demonstrating the increased yield at smaller \yt\ for the smaller $\eta$ acceptance. That result is consistent with the hard-component density in Fig.~\ref{etatcm} (right) assuming correspondence with dijets from low-$x$ partons (gluons) as the source for the spectrum hard component.
 }  
\end{figure} 

Figure~\ref{etacomp} shows inferred spectrum hard components in the form $H(y_t) / \bar \rho_s^2$ for two values of acceptance $\Delta \eta$. The dashed curves show fixed hard-component model function $\alpha \hat H_0(y_t)$ (Gaussian with power-law tail) for comparison. The densities $\bar \rho_0,~\bar \rho_s,~\bar \rho_h$ are derived within $\Delta \eta = 2$ for both cases. Above $y_t = 3$ ($p_t \approx 1.4$ GeV/c) the distributions are equivalent for all multiplicity classes.

Below $y_t = 3$ the \yt-spectrum hard component increases substantially within the smaller $\eta$ acceptance, nearer the peak of $\tilde H_0(\eta)$. The increase near $y_t = 2$ ($p_t \approx 0.5$ GeV/c) is a factor 2. That trend can be understood as follows: Near $\eta = 0$ an increased MB dijet fragment yield is expected at lower \yt\ because more lower-$x$ participant gluons are favored there, and the lowest-energy jets ($E_{jet} \approx 3$ GeV) may have FFs skewed to lower fragment momenta. At larger $\eta$ larger-$x$ partons are favored, and the higher-energy jets have harder FFs.

For the lowest multiplicity class the hard component falls mainly below $y_t = 2.7$ ($p_t \approx 1$ GeV/c) and may continue to increase below 0.5 GeV/c. That trend is consistent with the corresponding data in Fig.~\ref{etatcm} (right) that fall well below the trend for other multiplicity classes.


 \section{Systematic uncertainties} \label{syserr}

We consider uncertainties in TCM decomposition of charge multiplicity \nch\ and SP \yt\ spectra and $\eta$ densities. We demonstrate that the standard 2D fit model is a necessary and sufficient model for \pp\ angular correlations. And we consider uncertainties in jet-related and NJ quadrupole systematics resulting from 2D model fits.

\subsection{Integrated-multiplicity systematics}

Decomposition of corrected $n_{ch}$ within acceptance $\Delta \eta$ into soft and hard components $n_s$ and $n_h$ based on uncorrected yield $n_{ch}'$ relies on estimation of efficiency $\xi$ and coefficient $\alpha$ (Sec.~\ref{etayt}). Efficiency is determined via Eq.~(\ref{eff}) by integrating soft model $S_0'(y_t)$ that describes uncorrected spectrum data to obtain $\xi \approx 0.6$ reflecting a combination of low-\yt\ tracking inefficiency and \yt\ acceptance cutoff. The estimated value in Ref.~\cite{ppprd} was $\xi = 0.5$ with slightly higher \pt\ cutoff (0.20 vs 0.15 GeV/c).

The estimated value for $\alpha$ differs between \yt\ spectra and $\eta$ densities. In the former case $\alpha \approx 0.006$, in the latter $\alpha \approx 0.012$. The relevant value is within $\alpha \in 0.01\pm0.005$ but with significant systematic bias depending on the context. We speculate that the lower value for \yt\ spectra describes well the region $y_t > 2.7$ ($p_t > 1$ GeV/c) where the hard-component shape is the same for all \nch, but not the region below that point where the shape is strongly dependent on \nch\ and substantially exceeds $H_0(y_t)$. In contrast, the density on $\eta$ integrates over the entire \yt\ acceptance and includes those extra contributions, requiring a larger value for $\alpha$.

\subsection{Single-particle $\bf y_t$ spectra and $\bf \eta$ densities}

Definition of the soft component is a central issue for  TCM decomposition. The soft component is defined as the limiting case of ratio $\rho_0 / \bar \rho_s$ as $\bar \rho_s \rightarrow 0$. If Eq.~(\ref{ppspec}) or Eq.~(\ref{rho0eta2}) does describe corresponding data the fixed soft component $S_0(y_t)$ or $S_0(\eta)$ should emerge as a stable limiting case, and that is what we observe. Subtraction of the  soft-component model reveals a stable hard-component shape with exception of two low-\nch\ hard components where deviations occur mainly at lower \yt.

For the $\eta$ densities in Sec.~\ref{etadensity}, Fig.~\ref{etatcm} indicates that the TCM system is self-consistent at the level of statistical error with the exception of the lowest \nch\ class. That exception is likely due to bias toward lowest-energy jets near mid-rapidity (for small \nch\ condition) where the mean fragment distribution extends down into the inefficient low-\yt\ region as shown in Fig.~\ref{etacomp}. One indication of self-consistency is the opposite trends on $\eta$ of soft and hard components in Fig.~\ref{etatcm} suggesting minimal crosstalk.

Uncertainties in the TCM for \yt\ spectra are discussed at length in Ref.~\cite{ppprd}. Uncertainty in the soft component increases substantially below $y_t \approx 2.5$ ($p_t \approx 0.8$ GeV/c), but the hard component falls off sharply below that point, with apparent lower bound for \pp\ collisions near $y_t = 1.6$ ($p_t \approx 0.35$ GeV/c). Thus, the hard-component absolute magnitude near and above its mode is well-defined, and \nch\ dependence below that point relative to the fixed soft component remains informative.

\subsection{Necessary and sufficient 2D fit model}

Although the standard 2D fit model in Eq.~(\ref{modelfunc}) applied to \auau\ data provides an excellent overall description of those 2D histograms~\cite{anomalous} its uniqueness may be questioned. For instance, is a specific data component divided among two or more model elements (jet peak and ``ridge''), do multiple data components contribute to a single model element (``flow'' and ``nonflow''), are additional model elements required to describe some data (e.g.\ ``higher harmonics'')?

Those questions have been addressed in several studies. Arguments against ``higher harmonics'' based on extensive data analysis are presented in Refs.~\cite{multipoles,sextupole}. A comparison of several candidate 1D fit models based on Bayesian inference methods in Ref.~\cite{bayes} demonstrates that a 1D projection of the standard 2D fit model is overwhelmingly preferred over other candidates (e.g.\ Fourier series). A recent study demonstrates that 1D models based on Fourier series confuse contributions from two sources of azimuth quadrupole structure (``flow'' and ``nonflow'')~\cite{v2ptb}.

In the present study we apply the standard model for \aa\ collisions to \pp\ data although the quadrupole model element is not {\em required} by MB \pp\ correlation data ($A_Q$ is not significant for the lowest multiplicity class). However, the \nch\ trend from this study makes clear that a NJ quadrupole element is definitely required for larger \nch\ values. The standard 2D fit model is thus established as  {\em necessary} for these \pp\ data, and the fit residuals demonstrate that the standard model is {\em sufficient} (all fit residuals are consistent  with statistical errors, see Fig.~\ref{fits}).

\subsection{Jet-related systematics}

Four correlation components have substantial amplitudes near the angular origin: the SS 2D jet peak, the soft component, conversion-electron pairs and BEC. There is thus a potential near the origin for cross-talk among correlation components and model elements, especially since dijet and BEC pair numbers have the same quadratic trend on \nch. Charge combinations combined with correlation shapes and \pt\ dependence permit distinctions. 

The jet-related SS 2D peak corresponds mainly to US pairs (local charge conservation during parton fragmentation). BEC modeled by a 2D exponential relates only to LS pairs.  Conversion electron pairs relate only to US pairs but the peak is very narrow. The soft component (nearly uniform on $\phi_\Delta$) relates only to US pairs. Distinctions among correlation components based on their charge properties have been explored in Refs.~\cite{porter2,porter3}. 

Jet-related components are almost completely separated from soft and BEC/electron components respectively above and below $p_t \approx 0.5$ GeV/c (Fig.~\ref{ppcorr}). However, for \yt-integral studies (as in the present case) there may be some cross talk between SS 2D peak and soft component depending on their relative amplitudes. And the BEC/electron peak may contribute a small fixed fractional bias to the SS 2D jet peak independent of \nch.

The major covariances are between the SS peak $\phi$ width and the soft component (Fig.~\ref{ss2dpar}, right), and between soft and SS 2D peak amplitudes (Fig.~\ref{quadamp1}, right, last point). The AS 1D peak (AS dipole) is immune to SS peak covariances and may be used as a reference for dijet production.  The SS 2D peak volume and AS 1D amplitude should scale together with \nch\ reflecting the common $n_j$ dijet production trend, and that correlation is confirmed in Fig.~\ref{ppcent1}.

\subsection{NJ quadrupole systematics}

The relevance of a NJ quadrupole component for 200 GeV \pp\ collisions may be questioned, especially for the MB case. In modeling a jet-related AS 1D peak there is ambiguity between a 1D Gaussian or a combination of dipole and quadrupole multipoles (as in the standard 2D model). For a sufficiently broad AS peak the two models may be equivalent, and any quadrupole component inferred from the standard 2D model could be jet-related.

However, what we observe in Fig.~\ref{soft} (left) is equivalent to a cubic trend on $\rho_s$ for the NJ quadrupole pair number, distinct from the quadratic trend observed for jet-related spectrum and correlation components. And the cubic trend for the quadrupole in \pp\ collisions ($\propto N_{part} N_{bin}$) is comparable to the trend observed for the NJ quadrupole in \auau\ collisions ($\propto N_{part} N_{bin} \epsilon_{opt}^2$) where other factors argue for a unique NJ quadrupole component~\cite{v2ptb}. Thus, we conclude that the NJ quadrupole inferred for \pp\ collisions from the standard 2D model is significant and signals the presence of a unique quadrupole mechanism in elementary collisions. While such a mechanism is evidently ``collective'' in the sense that it involves the correlated motion of multiple FS hadrons, its interpretation in terms of hydrodynamic flow as a result of particle rescattering in a small transient system is questionable.

\section{LHC $\bf p$-$\bf p$ ``ridge" and SS curvatures} \label{ridgecms}

Three manifestations of so-called ``ridge'' phenomena reported at the RHIC and LHC can be distinguished: 
(i) $\eta$ elongation of a monolithic SS 2D jet peak in \aa\ collisions well described by a single 2D Gaussian~\cite{axialci,anomalous}, 
(ii) claimed development of a separate $\eta$-uniform ridge-like structure beneath a symmetric 2D jet peak also in \aa\ collisions~\cite{starridge}, and 
(iii) appearance of an SS ridge in \pp\ collisions for certain kinematic cut conditions~\cite{cms,cmsridge}. 
Item (i) is well established for untriggered (no $p_t$ cuts) jet correlations but has been referred to as a ``soft ridge"~\cite{glasma2,glasma}, although there is no separate ridge per se distinguished from the SS 2D jet peak. Item (ii) appears for certain combinations of $p_t$ cuts (``triggered'' jet analysis) and may also be jet-related.

We refer here to item (iii)---the so-called ``CMS ridge.'' Appearance of a SS ridge in CMS data for some cut combinations might be associated with item (ii) above, given the apparent similarity. However, the absence of $\eta$ elongation as in (i) and consistency with measured NJ quadrupole systematics makes interpretation (ii)  unlikely. There is no indication from this study that the CMS ridge is directly associated with the SS 2D jet peak. 

A study in Ref.~\cite{cmsridge} demonstrated that extrapolation of NJ quadrupole trends from \auau\ data to \nn\ (\pp) collisions and from RHIC to LHC energies could account quantitatively for the observed CMS ridge. One motivation for the present study has been confirmation of a significant NJ quadrupole amplitude in \pp\ collisions and determination of its \nch\ dependence. We now relate that new information to the CMS ridge phenomenon.

\subsection{Azimuth curvatures and the ``ridge'' phenomena} \label{curves}

Given the properties of 2D angular correlations from \pp\ collisions as reported in this study, structure such as the SS ridge reported by CMS for 7 TeV \pp\ data~\cite{cms} may result from competition between two curvatures on $\phi_\Delta$ within $|\eta_\Delta| > 1$ (which excludes most of the SS 2D jet peak). A SS ridge may appear when azimuth structure near $\phi_\Delta = 0$ in that region becomes significantly concave downward (negative {\em net} curvature). For MB \pp\ collisions only positive net curvatures appear, but a small {\em relative} change in certain correlation amplitudes may result in qualitative appearance or disappearance of a ridge.

As reported in Sec.~\ref{modelfits} the dominant structures within $|\eta_\Delta| > 1$ are the AS dipole sinusoid $\cos(\phi_\Delta - \pi)$ and the azimuth quadrupole sinusoid $\cos(2\phi_\Delta)$. At a point where the slope of a function $f(x)$ is zero its curvature $k$ is just the second derivative $k = f''(x)$. The curvature at a maximum is negative. The curvatures of the cylindrical multipoles $\cos[m(\phi - \pi)]$ at $\phi = 0$ are $(-1)^{m+1} m^2$. The curvature of $\cos(2\phi_\Delta)$ at $\phi_\Delta = 0$ is then four times the curvature of $\cos(\phi_\Delta - \pi)$ and with opposite sign. 

Absolute curvatures are determined by the coefficients of the two sinusoids -- $A_D / 2$ for dipole and $2A_Q$ for quadrupole. In defining the separate curvatures we include  a common factor $\bar \rho_0 / \bar \rho_s$ to take advantage of the simple trends in Figs.~\ref{ppcent1} (right) and \ref{soft} (left).
The curvatures for dipole and quadrupole respectively are then $k_D = (\bar \rho_0 / \bar \rho_s) A_D/2$ and $k_Q =-8  (\bar \rho_0 / \bar \rho_s) A_Q$.  Zero net curvature corresponds to $ 4 \times 2 A_Q = A_D / 2$ or $-k_Q / k_D = 16 A_Q / A_D = 1$. A ridge (negative net curvature) may be identified if that ratio is significantly greater than one.


 \begin{figure}[h]
  \includegraphics[width=3.3in,height=1.6in]{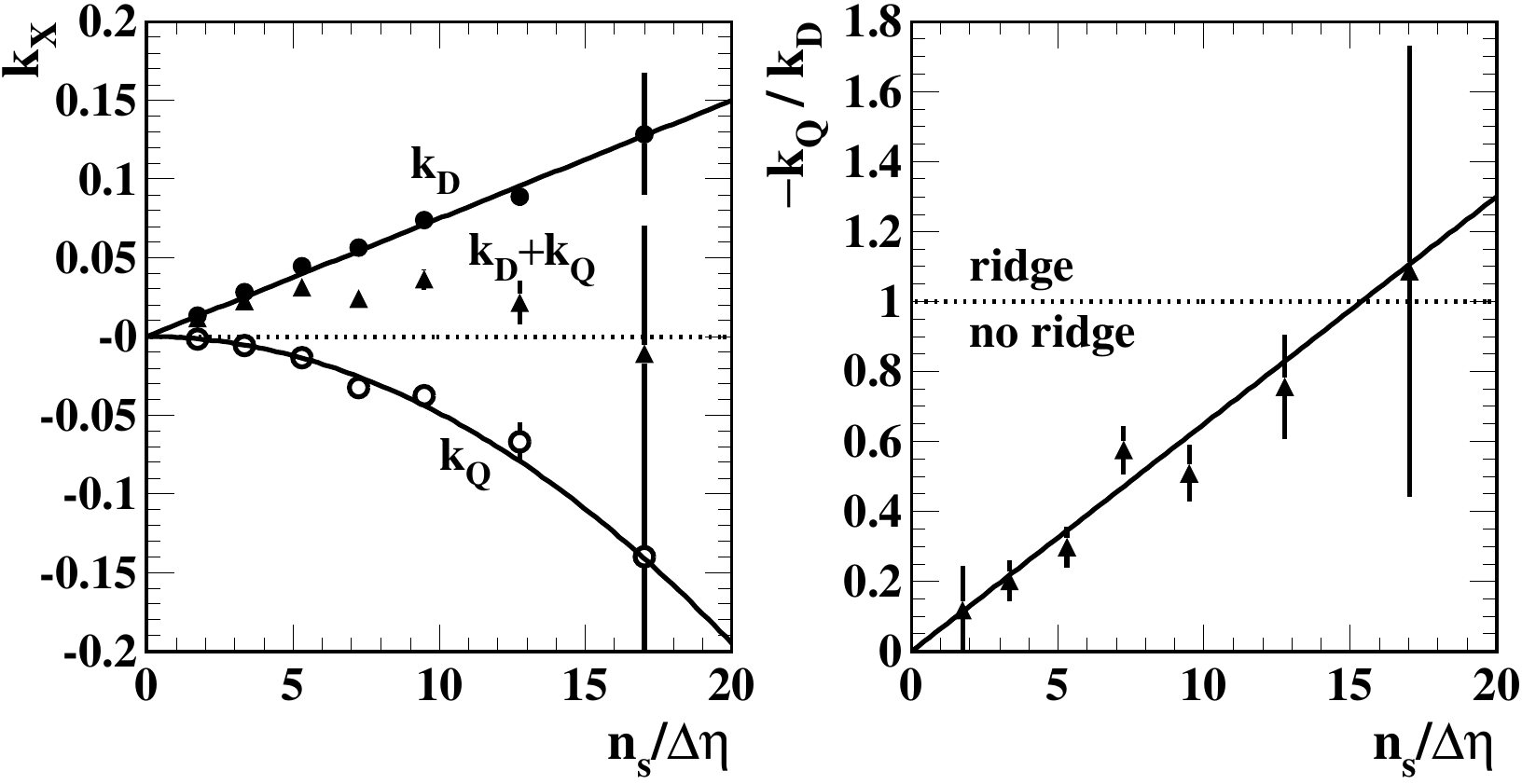}
\caption{\label{curvature}
Left: Azimuth curvatures $k_D$ and $k_Q$ for dipole and quadrupole components respectively and net curvature $k_D + k_Q$, all evaluated at $\phi_\Delta = 0$.
Right: Curvature ratio $-k_Q / k_D$ that predicts when a SS ``ridge'' will appear within $|\eta_\Delta| > 1$ at $\phi_\Delta = 0$.
 }  
 \end{figure}

Figure~\ref{curvature} (left) shows curvatures $k_X$ at $\phi_\Delta = 0$ vs soft density $\bar \rho_s$ (proxy for number of participant low-$x$ gluons). The net curvature $k_D + k_Q$ (solid triangles) is also shown. As determined by the trends in Figs.~\ref{ppcent1} (right) and \ref{soft} (left) the dipole curvature is positive and increases linearly while the quadrupole curvature is negative and increases (in magnitude) quadratically. The net curvature is mainly positive but consistent with zero for the highest multiplicity class, although the fit error is large.

Figure~\ref{curvature} (right) shows the ratio $-k_Q / k_D = 16 A_Q / A_D$ vs $\bar \rho_s$ for 200 GeV \pp\ collisions increasing linearly (solid line) from zero and passing through unity near  the highest multiplicity class. As noted, for ratios significantly greater than 1 (significant negative net curvature) a SS ridge should appear within $|\eta_\Delta| > 1$. At higher collision energies and with additional \pt\ cuts the slope of the linear trend should increase, resulting in appearance of a significant SS ridge even for modest charge densities.

 \begin{figure}[h]
\includegraphics[width=1.6in]{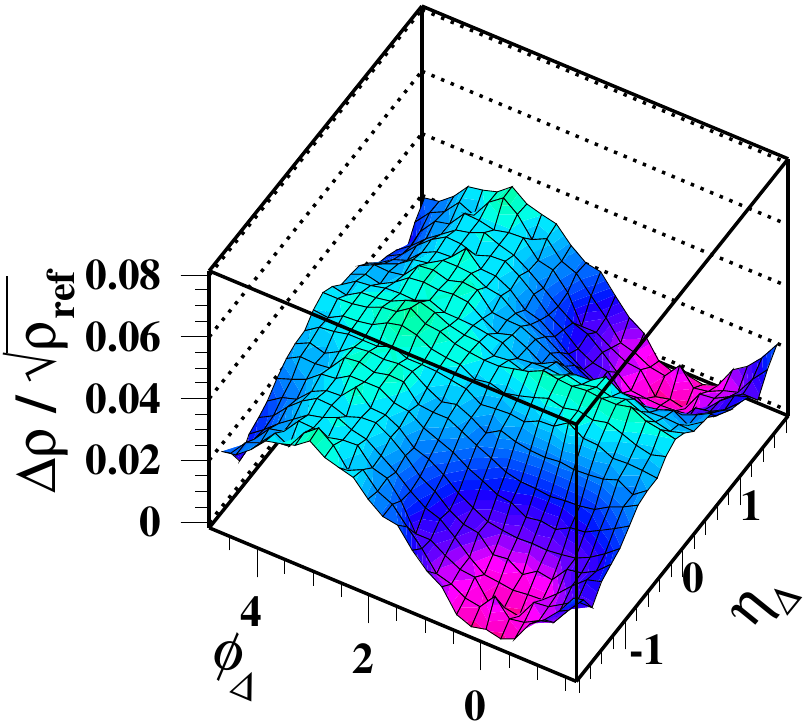}
\includegraphics[width=1.6in]{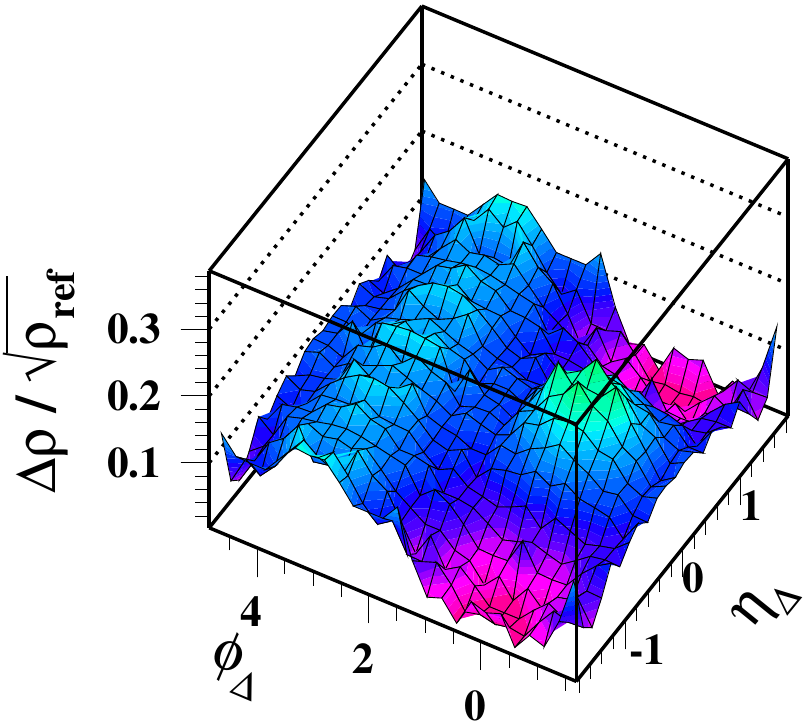}
\caption{\label{quadcomp}
2D angular correlations for  n = 1 (left) and 6 (right) multiplicity classes from 200 GeV \pp\ collisions. Fitted model elements for soft component, BE + conversion electrons and constant offset have been subtracted from the data leaving jet-related and NJ quadrupole data components.
}  
 \end{figure}

Figure~\ref{quadcomp} shows 2D angular correlations for $n = 1$ (left) and 6 (right) multiplicity classes. Fit-model elements for the soft component, BE + conversion electrons and constant offset have been subtracted from the data leaving the jet-related and NJ quadrupole components (see Fig.~\ref{fits} for fit details). In the region $|\eta_\Delta| > 1$ that excludes the SS 2D peak we observe that  near $\phi_\Delta = 0$ (SS) the curvature varies from large and positive (left) to consistent with zero (right). Near  $\phi_\Delta = \pi$ (AS) the negative curvature {\em approximately doubles}. The second variation is as much a part of the ``ridge'' phenomenon as the first, but is typically not acknowledged~\cite{cms,cmsridge}.

\begin{table*}[t]
\caption{ \label{table}
\pp\ correlation systematics. Entries represent systematics inferred from RHIC data at 0.2 TeV and extrapolations inferred by modeling CMS angular correlation histograms at 7 TeV from Ref.~\cite{cms}.  The left columns indicate collision energy and cut conditions, including minimum-bias (MB) data. The $A_X$ represent parameters from the present per-particle analysis with an intensive correlation measure. 
The corresponding CMS measures are defined by $R_x = (2\pi \times 4.8) A_X \approx 30 A_X$~\cite{cmsridge}.
} 
\begin{tabular}{|c|c|c|c|c|c|} \hline
 $\sqrt{s}$ (TeV) & Condition & $100A_{Q}$    & $A_D$  & $A_{2D} $  & $16 A_Q / A_D$    \\ \hline
 0.2  & NSD & $0.05\pm 0.02$ & 0.05$\pm0.002$ & 0.06 $\pm0.002$ &  0.16$\pm0.07$  \\ \hline 
 0.2  & $n = 6$ & $0.84\pm 0.15$ & 0.18$\pm 0.012$ & 0.26$\pm0.05$ &  0.75$\pm0.17$  \\ \hline 
 7  & MB & $0.12\pm0.04$ & 0.07$\pm0.01$  & 0.14$\pm0.01$ &  0.27$\pm0.12$  \\ \hline 
 7  &  $N_{trk} > 110  $ & $1.8\pm0.6$ & 0.28$\pm0.04$  & 0.56$\pm0.04$ &  1.03$\pm0.3$ \\ \hline 
 7  & $p_t$,  $n_{ch}  $  cuts & $1.2\pm0.4$ & 0.14$\pm0.02$  & 0.28$\pm0.02$ & 1.37$\pm0.4$   \\ \hline 
 \end{tabular}
\end{table*}

Alteration of the SS 2D peak is also notable. Visually the peak appears to narrow dramatically on $\phi_\Delta$ with increasing \nch. However, 2D model fits reveal that the fitted peak width decreases only slightly. The {\em apparent} narrowing is due to superposition of the NJ quadrupole component onto the 2D peak structure. The change from left to right panel is dominated by the ten-fold increase of curvature ratio $|k_Q / k_D|$, as indicated in Fig.~\ref{curvature} (right).

\subsection{RHIC vs LHC correlation structure}

In Ref.~\cite{cmsridge} the systematics of 2D angular correlations derived from 62 and 200 GeV \auau\ collisions were extrapolated first to peripheral \nn\ collisions (as a proxy for NSD \pp\ collisions) and then to 7 TeV for comparisons with CMS \pp\ data.  The question posed: Are SS 2D peak, AS dipole and NJ quadrupole systematics at and below 200 GeV consistent  with those at 7 TeV and especially with the emergence of a SS ``ridge'' structure for certain conditions imposed at that energy? As to \pp\ estimates, at 200 GeV the NSD \pp\ values of $A_{2D}$ and $A_D$ from the present study are numerically consistent with the \auau\ extrapolation to \nn\ collisions. The value of $A_Q$ for 200 GeV \pp\ collisions was overestimated by a factor 2 by addition of a conjectured quadrupole contribution from the AS 1D peak modeled as a 1D Gaussian. According to the present study the AS peak for \pp\ collisions is actually well described by a single dipole element.

It was demonstrated that dijet production at 7 TeV is consistent with extrapolation from 200 GeV using  factor $R(\sqrt{s}) = 2.3$ (derived from comparison of 62 and 200 GeV data) interpreted to describe the increase of participant low-$x$ gluons with increasing collision energy. That factor applies to the {\em per-particle} SS 2D peak amplitude $A_{2D}$ (intrajet correlations), whereas increase of AS dipole amplitude $A_D$ (jet-jet correlations) is considerably less (consistent with no amplitude increase from 62 to 200 GeV~\cite{anomalous}.  NJ quadrupole measurements at RHIC suggest that $A_Q$ also increases by factor 2.3.

As to multiplicity trends for \pp\ collisions, in the present study the corrected charge density for multiplicity class $n = 6$ is $\bar \rho_0 = 13.7$, 5.5 times the MB value 2.5, whereas at 7 TeV the CMS $N_{trk} > 110$ multiplicity class corresponds to corrected $\bar \rho_0 = 136 / 4.8 = 28$, 4.8 times the MB value 5.8. In this study the $n = 6$ class corresponds to measured four-fold increase of $A_D$ and $A_{2D}$ and fifteen-fold increase of $A_Q$ over their MB values.

As to responses to \pt\ cuts we note that about half of all MB jet fragments appear below the mode of the 200 GeV spectrum hard component at 1 GeV/c~\cite{ppprd,fragevo}. In contrast, the mode of $v_2(p_t)$ on \pt\ is close to 3 GeV/c~\cite{2004}. The CMS cut $p_t \in [1,3]$ GeV/c is effectively a lower limit at 1 GeV/c, since the hadron spectrum falls rapidly with \pt. We thus expect that a \pt\ lower limit imposed at 1 GeV/c should reduce the AS dipole substantially more than the NJ quadrupole. We have reduced $A_D$ by factor 1/2 and $A_Q$ by factor 2/3 to estimate the effect of \pt\ cuts.

Table~\ref{table} summarizes the various estimates. The first two rows report results from the present study and the curvature ratios shown in Fig.~\ref{curvature}. As noted in the text the 7 TeV MB values are obtained by multiplying $A_{2D}$ and $A_Q$ by 2.3 and $A_D$ by 1.4. The values for $N_{trk}>110$ are obtained with factors 4 and 15 applied as for the 200 GeV values for $n = 6$ (ignoring the small difference in ratios to MB multiplicities between RHIC and LHC energies). The effect of the \pt\ cut is estimated by factors 1/2 and 2/3 as noted above.

Results from the present study describe the reported CMS 7 TeV 2D angular correlations quantitatively and are generally consistent with Ref.~\cite{cmsridge} but also provide insight into the physical origins of the reported SS ``ridge.'' The large collision-energy increase combined with imposed $p_t$ and multiplicity cuts increases the \pp\ NJ quadrupole amplitude eight-fold relative to the AS 1D jet peak, changing the SS curvature sign and producing an apparent SS ridge. In effect, the SS azimuth curvature functions as a comparator, switching from valley to ridge as one amplitude increases relative to another. A quantitative curvature change is  transformed to a qualitative shape change (mis)interpreted as emergence of a novel phenomenon at a higher energy.

 \section{Discussion} \label{disc}

Several open issues for high-energy \pp\ collisions were summarized in Sec.~\ref{issues}: (a) the role of \pp\ collision centrality,   (b) the definition and nature of the \pp\ underlying event or UE, (c) the systematics of MB dijet production and (d) confirmed existence and characteristics of a NJ quadrupole component in \pp\ angular correlations. We  return to those points in light of results from this study.

\subsection{Dijets vs NJ quadrupole vs flows}

The measured hard components of \yt\ spectra, $\eta$ densities and 2D angular correlations presented in this study complete a unified experimental and theoretical picture of MB dijet production (no \pt\ cuts) established previously for \auau\ collisions~\cite{anomalous,hardspec,fragevo}. There were no previous measurement of a NJ quadrupole component for \pp\ collisions. The combined dijet and quadrupole results from the present study convey important implications for claims of \pp\ collectivity (flows), \pp\ centrality, UE studies and the mechanism of the CMS ridge.

The term ``collective'' or ``collectivity'' (e.g.\ as recently applied to small collision systems at the LHC) is ambiguous, since jet formation is a form of ``collectivity'' as is the NJ quadrupole whatever its production mechanism. Introducing the term ``collectivity'' as synonymous with ``flow'' may produce confusion. There are certainly collective aspects of \pp\ collisions although it is unlikely that hydrodynamic flow (in the sense of fluid motion resulting from particle rescattering) plays a role. Dijet production and the NJ quadrupole amplitude follow characteristic trends on $\bar \rho_s$ precisely over a large range of amplitudes (100-fold for dijets, 1000-fold for quadrupole as correlated-pair numbers) while the underlying particle (participant-gluon) density varies 10-fold. How can a small collision system with extremely low particle density support a hydrodynamic phenomenon that conspires to follow the same trends over such a large density interval?

\subsection{$\bf p$-$\bf p$ centrality and IS  geometry}

The notion of centrality (impact parameter) for \pp\ collisions is ambiguous in principle. Concerning \pp\ centrality several questions arise: What does ``IS geometry'' mean for \pp\ collisions? Is an impact parameter relevant? How are total \nch, triggered dijets, transverse low-$x$ parton (gluon) density and \pp\ centrality correlated? How do those factors relate to measured ensemble-mean PDFs, {\em event-wise} participant-parton distributions on $x$ and initial momentum transfer?
There are certainly large event-wise fluctuations in soft-hadron and dijet production, but whether those correspond to fluctuations of \pp\ IS transverse geometry or some other collision aspect is in question. The need for comprehensive study of the \nch\ dependence of \pp\ angular correlations in relation to \pp\ centrality was one motivation for the present study. 

Results from this study support two arguments against a major role for a \pp\ centrality concept: (a) \pp\ dijet production scales as $N_{bin} \propto N_{part}^2 \sim \bar \rho_s^2$  but the eikonal approximation implies binary-collision scaling as  $N_{bin}(b) \propto N_{part}^{4/3}(b)$ (as for \aa\ collisions). The observed dijet trend is consistent with encounters between all possible participant-gluon pairs in each collision, not a smaller subset determined by an impact parameter. (b) The NJ quadrupole amplitude scales as $N_{part} N_{bin} \sim \bar \rho_s^3$ over a large \nch\ range consistent with {\em part} of the $N_{part}N_{bin} \epsilon_{opt}^2$  trend observed for \aa\ collisions,  but there is no significant reduction with a decreasing \pp\ eccentricity. The combined trends suggest that \pp\ IS geometry is not a determining factor for either phenomenon. Instead, the event-wise depth of penetration on momentum fraction $x$ of the projectile wave function may be the main source of variation for soft, hard and quadrupole components.


\subsection{Implications for UE studies}

As noted in Sec.~\ref{issues} UE studies rely on several assumptions: (a) Concentration of low-$x$ gluons at small radius in the proton (inferred from DIS data) provides a correlation among \pp\ centrality, soft hadron production and dijet production, (b) the conventionally-defined TR on azimuth includes no contribution from a triggered dijet and (c) multiple-parton interactions (MPI) may occur in jet-triggered \pp\ collisions. The integrated TR yield denoted by $N_\perp$ is observed to increase with increasing jet-trigger \pt\ condition and is interpreted to represent a soft background increasing with \pp\ centrality.

Reference~\cite{pptheory} addressed part of that narrative with simulations based on the TCM for hadron production from \pp\ collisions. It concluded that most dijets (what would result from an applied trigger at lower \pt) emerge from {\em low}-multiplicity collisions. Spectrum studies already indicated that higher-multiplicity collisions do produce dijets at higher rates but are few in number and so contribute only a small fraction of the \pt-triggered event population. A $p_{t}$ condition  cannot significantly alter the soft component or \pp\ collision centrality (if relevant). 

The present study adds the following new information: (a) 2D angular correlations confirm a strong contribution from MB dijets {\em within the TR}. (b) The dijet production trend $\propto \bar \rho_s^2$ suggests that the eikonal approximation is invalid and that centrality is not a useful concept for \pp\ collisions. (c) Monotonic increase of the NJ quadrupole $\propto \bar \rho_s^3$ over a large \nch\ range also suggests that \pp\ centrality, as manifested in a varying IS eccentricity, is not a useful concept. Those factors confirm the conclusions of Ref.~\cite{pptheory} and lead to the following scenario for $N_\perp$ variation with a \pt\ trigger: As the trigger condition is increased from zero the integrated spectrum soft component increases $N_\perp$ from zero to a plateau on $p_{t,trig}$. The hard-component (jet) contribution to $N_\perp$ is similarly integrated up to a plateau. $N_\perp$ thus has both soft and hard components exhibiting plateau structures slightly displaced from one another on $p_{t,trig}$. Almost all events satisfying an increased trigger condition contain a dijet (are hard events) but remain characteristic of a MB population with smaller soft multiplicity, not the expected more-central population with larger soft multiplicity.

Given the observed \nch-dependent structure of \pp\ 2D angular correlations we arrive at three conclusions: (a) All dijets include a large-angle base that strongly overlaps the TR. That base dominates minimum-bias jets but may persist within all higher-energy (e.g.\ \pt-triggered) dijets. (b) The region with a minimal dijet contribution that might suffice for UE studies is defined by $|\eta_\Delta| > 1$ and $\phi_\Delta \approx 0$ (see Fig.~\ref{quadcomp}). An immediate example of novel UE structure that might be discovered there is provided by the CMS ``ridge,''  a manifestation of the NJ quadrupole that was not expected in \pp\ collisions. (c) The likelihood of multiple dijet production in \pt-triggered events (which retain a low soft multiplicity as noted) is small whereas the likelihood of multiple dijets in high-multiplicity events approaches unity.  The usual interpretation of $N_\perp$ trends in terms of MPI may be misleading.

\subsection{TCM confirmation and self-consistency}

Arguments against the TCM have been presented since commencement of RHIC operation. It has been noted that the HIJING Monte Carlo~\cite{minijets} (based on PYTHIA~\cite{pythia}) fails to describe RHIC and LHC \aa\ data. That failure as been expressed as ``too slow increase'' of hadron production with centrality and energy~\cite{global2,perfect}. HIJING is assumed to represent the TCM and its failure is then confused with failure of the TCM itself, of which HIJING is only a specific theory implementation. The problems with HIJING are traceable to the eikonal-model assumption included in default PYTHIA~\cite{anomalous}. Such arguments typically rely on data from a centrality range covering only the more-central 40-50\% of the \aa\ cross section~\cite{cgcmult,perfect}. The critical centrality region extending from \pp\ or \nn\ collisions to the {\em sharp transition} in jet formation~\cite{anomalous} is not considered. Alternative models that seem to describe the more-central data are actually falsified by more-peripheral data~\cite{glasma}. An alternative argument is based on assuming that TCM agreement with data is accidental, that a {\em constituent-quark} model (soft only, excluding jets) is more fundamental and describes data as well~\cite{phenix}, but that argument is questionable~\cite{nominijets}.

The TCM invoked in this study is based on previous analysis of \pt\ spectrum $n_{ch}$ and centrality dependence~\cite{ppprd, anomalous,hardspec,fragevo,tomalicempt}, \pt\ fluctuations and correlations~\cite{ptscale,ptedep,tomaliceptfluct}, transverse-rapidity $y_t \times y_t$ correlations~\cite{porter2,ytxyt} and minimum-bias 2D number angular correlations~\cite{porter2,porter3,anomalous}.  In each case model elements were determined quantitatively by systematic analysis without regard to physical mechanisms. Only after the TCM was so established were connections with theory and physical interpretations introduced.
In the present study we extend the TCM to describe the \nch\ dependence of 200 GeV \pp\ $\eta$ densities and 2D angular correlations. In the latter we observe for the first time a significant NJ quadrupole component and its \nch\ dependence as a novel nonjet phenomenon within the UE. We find  that the extended TCM remains fully self-consistent and provides accurate and efficient representation of a large body of data.

\section{Summary} \label{summ}

We report measurements of the charge-multiplicity dependence of single-particle (SP) densities on transverse rapidity \yt\ (as \yt\ spectra) and pseudorapidity $\eta$ and 2D angular correlations on $(\eta,\phi)$ from 200 GeV \pp\ collisions. The SP densities are described accurately by a two-component (soft + hard) model (TCM) of hadron production. The inferred \yt-spectrum TCM is consistent with a previous study. The  result for $\eta$ densities newly reveals the distribution on $\eta$ of minimum-bias (MB) jet fragments. 2D angular correlations are fitted with a multi-element fit model previously applied to data from 62 and 200 GeV \auau\ collisions. Fit residuals are consistent with statistical uncertainties in all cases.

Trends for several 2D correlation model parameters are simply expressed in terms of TCM soft-component multiplicity $n_s$ or mean density $\bar \rho_s \equiv n_s / \Delta \eta$ ($\Delta \eta$ is a detector acceptance). Correlated-pair numbers for soft component (projectile dissociation) scale $\propto \bar \rho_s$, for hard component (dijet production) scale  $\propto \bar \rho_s^2$ and for nonjet (NJ) quadrupole scale $\propto \bar \rho_s^3$. The NJ quadrupole amplitude is quite significant for higher-multiplicity \pp\ collisions.

The dijet production trend is inconsistent with an eikonal approximation for \pp\ collisions (which would require dijets $\propto \bar \rho_s^{4/3}$), and the monotonically-increasing NJ quadrupole trend is inconsistent with an initial-state eccentricity determined by \pp\ impact parameter. The two trends combined suggest that centrality is not a useful concept for \pp\ collisions. Fluctuations may instead depend on the event-wise depth of penetration on momentum fraction $x$ of the projectile wave functions and in turn on the number of participant low-$x$ partons.

 2D angular-correlation data are in conflict with assumptions relating to the \pp\ underlying event (UE, the complement to a triggered dijet). The azimuth transverse region (TR) bracketing $\pi / 2$ is assumed to contain no contribution from a triggered dijet, but minimum-bias dijets are observed to make a strong contribution there. The region with minimal jet contribution is defined by $|\eta_\Delta| > 1$ near the azimuth origin that excludes the same-side 2D jet peak and most of the away-side 1D jet peak.

The presence of a significant NJ quadrupole component and its multiplicity trend have several implications: (a)  initial-state transverse geometry does not appear to be a useful concept for \pp\ collisions as noted above, (b) the appearance of a NJ quadrupole component in a small system with negligible particle density contradicts the concept of a hydro phenomenon based on particle rescattering and large energy/particle density gradients and (c) the same-side ``ridge'' observed in \pp\ collisions at the large hadron collider (LHC), interpreted by some to suggest ``collectivity'' (flows) in small systems, results from an interplay of the jet-related away-side 1D peak and the NJ quadrupole that together determine the curvature on azimuth near the origin. When that curvature transitions from positive to negative (depending on collision energy and other applied cuts) a same-side ``ridge'' appears. 

In a hydro narrative the NJ quadrupole component interpreted as elliptic flow should represent azimuth modulation of radial flow detected as a modification of SP \yt\ spectra. But no corresponding modification is observed in \pp\ \yt\ spectra despite precise differential analysis.

This material is based upon work supported by the U.S.\ Department of Energy Office of Science, Office of Nuclear Physics under Award Number DE-FG02-97ER41020.

\end{document}